\documentclass[5p,final,authoryear,times]{elsarticle}
\journal{Journal of Neuroscience Methods}
\usepackage{graphicx}
\usepackage{amsmath}
\usepackage{amssymb}
\usepackage{calrsfs}
\usepackage{amsfonts}
\usepackage{mathrsfs}
\usepackage{float} 
\usepackage{soul}
\usepackage[pdftex,breaklinks,colorlinks,
linkcolor=blue,
anchorcolor=blue,
citecolor=blue,
urlcolor=blue,]{hyperref}
\usepackage{lscape}
\usepackage{marginnote}

\usepackage{courier}

\usepackage{xcolor}
\usepackage{xspace}

\newcommand{\mpar}[1]{}

\newcommand{\mnote}[1]{}
\newcommand{\rem}[1]{}

\newcommand{\new}[1]{{\color{black} #1}}

\renewcommand{\eqref}[1]{\textup{{Eq.~(\ref{#1})}}}
\newcommand{\figref}[1]{\textup{{Fig.~\ref{#1}}}}
\newcommand{\secref}[1]{\textup{{Section~\ref{#1}}}}
\newcommand{\tabref}[1]{\textup{{Table~\ref{#1}}}}

\usepackage{caption}
\makeatletter
\def\ps@pprintTitle{%
   \let\@oddhead\@empty
   \let\@evenhead\@empty
   \def\@oddfoot{\reset@font\hfil\thepage\hfil}
   \let\@evenfoot\@oddfoot
}
\makeatother

\begin{document}


\title{Realistic Microstructure Simulator (RMS): Monte Carlo simulations of diffusion in three-dimensional cell segmentations of microscopy images}

\author[cbi,cai2r]{Hong-Hsi Lee\corref{cor}}
\author[cbi,cai2r]{Els Fieremans}
\author[cbi,cai2r]{Dmitry S. Novikov}
\cortext[cor]{Corresponding author: Honghsi.Lee@nyulangone.org}
\address[cbi]{Center for Biomedical Imaging, Department of Radiology, New York University School of Medicine, New York, NY, USA}
\address[cai2r]{Center for Advanced Imaging Innovation and Research (CAI2R), New York University School of Medicine, New York, NY, USA}
\date{\today}

\begin{abstract}

\noindent\textit{Background:} Monte Carlo simulations of diffusion are commonly used as a model validation tool as they are especially suitable for generating the diffusion MRI signal in complicated tissue microgeometries.


\noindent\textit{New method:} Here we describe the details of implementing Monte Carlo simulations in three-dimensional (3{\it d}) voxelized segmentations of cells in microscopy images. Using the concept of the corner reflector, we largely reduce the computational load of simulating diffusion within and exchange between multiple cells. Precision is further achieved by GPU-based parallel computations. 

\noindent\textit{Results:} Our simulation of diffusion in white matter axons segmented from a mouse brain demonstrates its value in validating biophysical models. Furthermore, we provide the theoretical background for implementing a discretized diffusion process, and consider the finite-step effects of the particle-membrane reflection and permeation events, needed for efficient simulation of interactions with irregular boundaries, spatially variable diffusion coefficient, and exchange.

\noindent\textit{Comparison with existing methods:} To our knowledge, this is the first Monte Carlo pipeline for MR signal simulations in a substrate composed of numerous realistic cells, accounting for their permeable and irregularly-shaped membranes.

\noindent\textit{Conclusions:} The proposed RMS pipeline makes it possible to achieve fast and accurate simulations of diffusion in realistic tissue microgeometry, as well as the interplay with other MR contrasts. 
Presently, RMS focuses on simulations of diffusion, exchange, and $T_1$ and $T_2$ NMR relaxation in static tissues, with a possibility to straightforwardly account for susceptibility-induced $T_2^*$ effects and flow. 

\end{abstract}


\maketitle

\section{Introduction}
The MRI measurements of self-diffusion of water molecules in biological tissues provide the sensitivity to the diffusion length scales ranging from microns to  tens of microns at clinically feasible diffusion times. As the feasible range of diffusion lengths is commensurate with the sizes of cells, diffusion MRI allows one to evaluate pathological changes in tissue microstructure \mpar{R1.10}\new{in vivo}. To balance between accuracy and precision in estimation of tissue parameters through biophysical modeling of diffusion MR signal, assumptions are inevitably made to simplify tissue microgeometry \citep{grebenkov2007review,jones2010book,kiselev2017mrphysics,Jelescu2017,Alexander2018,novikov2019review}. 
It is necessary to validate the assumptions of models before use, 
either through experiments in physical phantoms \citep{fieremans2018cookbook}, 
or testing the model functional forms in animals and human subjects \citep{novikov2018onmodeling}, 
or numerical simulations \citep{fieremans2018cookbook}.

\begin{figure*}[th!]
\centering
	\includegraphics[width=0.8\textwidth]{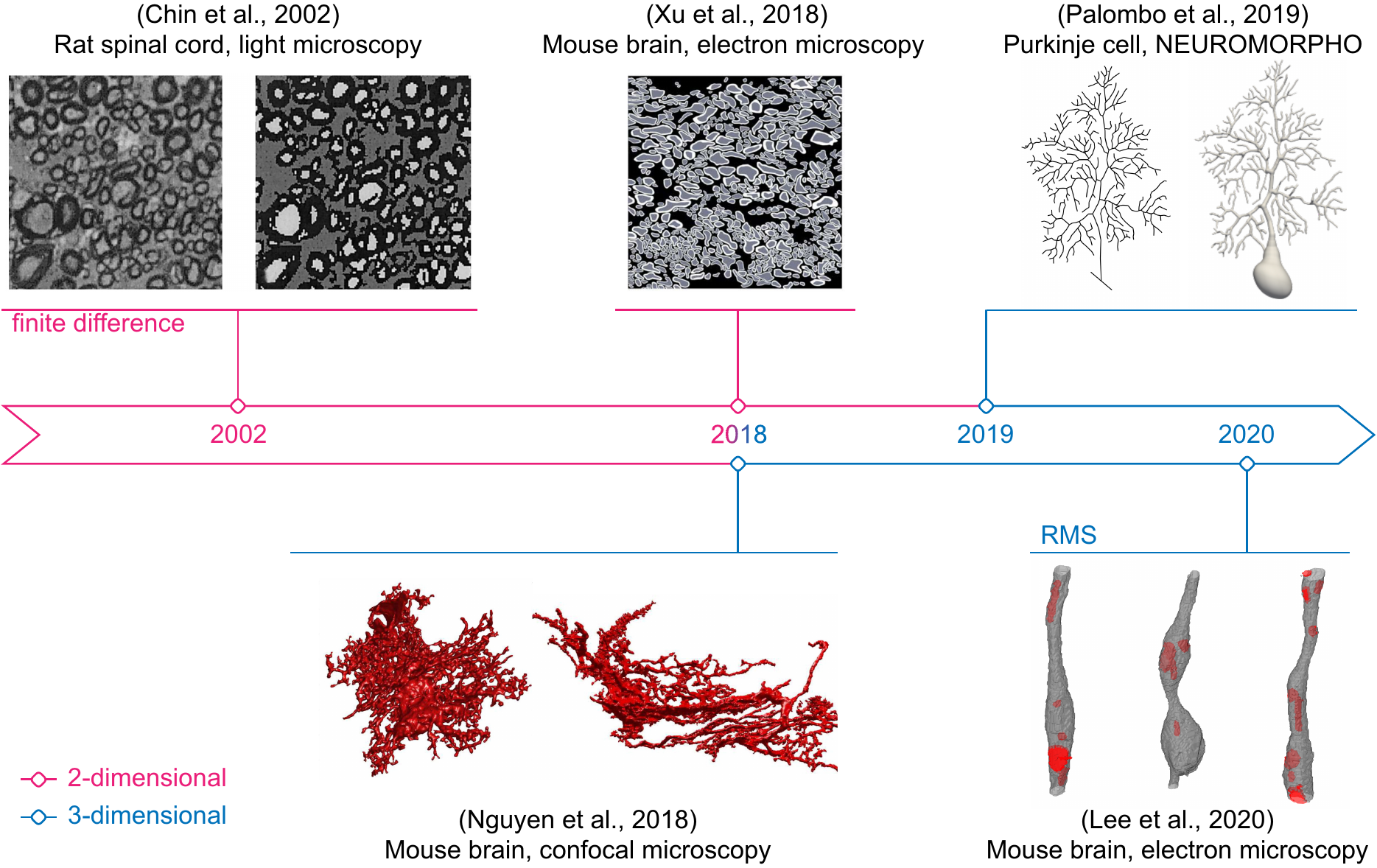}
	\caption{\textbf{Benefiting from advances in microscopy techniques and from increases in computational power, the research focus of diffusion simulations has gradually switched to 3-dimensional tissue microgeometries.} Realistic white matter axons in \citep{lee2020axial} are examples of voxelized simulation substrates used in RMS. The figure is adapted from \citep{chin2002simulation,xu2018susceptibility,nguyen2018gpu,palombo2019astrocyte,lee2020axial} with permission from Wiley, Elsevier, and Springer Nature.
	}
	\label{fig:history}
\end{figure*}

So far, numerical simulation is the most flexible and economic choice among all kinds of validation. Benefiting from the recent advances in microscopy, realistic cell geometries for simulations have been directly reconstructed from the microscopy data of neuronal tissues in 2 dimension (2{\it d}) \citep{chin2002simulation,xu2018susceptibility} and 3{\it d} \citep{nguyen2018gpu,palombo2019astrocyte,lee2020axial,lee2020radial}, as shown in \figref{fig:history}. The emerging need for simulations in realistic substrates prompts the development of open-source software congenial to physicists, biologists and clinicians.

Here, we describe our implementation of Monte Carlo (MC)-based diffusion simulations: the Realistic Microstructure Simulator (RMS), which entails a fast and accurate model validation pipeline. 
While our pipeline has been recently announced and applied to simulate diffusion MRI in axonal microstructure \citep{lee2020axial,lee2020radial,lee2020mchuman}, 
these publications are mainly focused on the physics of diffusion and model validation. In this work, we describe the methodology in detail, building on the algorithms introduced by our team over the past decade \citep{fieremans2008simulation,fieremans2010karger,novikov2011rpbm,novikov2014meso,burcaw2015meso,fieremans2018cookbook,lee2020gray}, and in particular, derive the finite MC-step effects relevant for the interactions (reflection and permeation) of random walkers and  membranes. 

RMS is introduced as follows.
In \secref{sec:methods}, we provide an overview of RMS implementation. 
Theoretical results and implementation details of particle-membrane collisions and exchange are presented in \ref{sec:app-erl} and \ref{sec:app-rs}; 
the first order correction of membrane permeability due to a discretized diffusion process is derived in \ref{sec:app-permeability}.  
In \secref{sec:results}, we demonstrate the application of RMS to diffusion simulations in realistic axonal shapes reconstructed from electron microscopy data of a mouse brain \citep{lee2019em}. 
Simulated diffusion MR signals are shown to be closely related to features of cell shape, facilitating the interpretation of diffusion measurements in biological tissues. 
Finally, in \secref{sec:discussion} we provide an outlook for microstructure simulation tools in general, and RMS in particular, as a platform for MR-relevant simulations of diffusion and relaxation in microscopy-based realistic geometries.  


\section{RMS Implementation}
\label{sec:methods}

\subsection{Realistic Microstructure Simulator: An overview}
\label{sec:rms-general}
The goal of our RMS implementation is to provide a universal platform of MC simulations of diffusion in any realistic microgeometry based on microscopy data. Therefore, the RMS has the following properties:
\begin{enumerate}[(i)]
    \item The simulation is performed in 3{\it d} continuous space with voxelized microgeometry. We will introduce this main feature of RMS in 
    \secref{sec:substrates}.
    \item The particle-membrane interaction of impermeable membrane is modeled as classic elastic collision. The reason for this choice will be explained in \secref{sec:interaction}, \ref{sec:app-erl} and \ref{sec:app-rs}.
    \item The models of permeable and absorbing membranes are implemented based on the novel theoretical results tailored for diffusion simulations of high accuracy, with details in \secref{sec:membrane-permeation} and \ref{sec:app-permeability}.
    \item The interplay of diffusion and other MR contrasts, such as $T_1$ and $T_2$ relaxation, surface relaxation and magnetization transfer effects, are incorporated in MC simulated Brownian paths and MR signal generation (\secref{sec:contrast}).
    \item The simulation kernel is accelerated by parallel computation on the GPU (\secref{sec:gpu}).
\end{enumerate}

\subsubsection{Substrates and masks}
\label{sec:substrates}

To generate the 3{\it d} substrates based on microscopy data for diffusion simulations, we can translate the voxelized cell segmentation into either smoothed meshes or binary masks \citep{nguyen2018gpu}. Each approach has its own pros and cons.

For the smoothed-mesh approach, the generated cell model has smooth surface, potentially having surface-to-volume ratio similar to the real cells. However, it is non-trivial to decide on the degree of smoothing while generating the cell model. In addition, in simulations, the problem of floating-point precision may arise, especially for determining whether a random walker encounters a membrane.

For the binary mask approach, it is fast and simple to translate the discrete microscopy data into the 2{\it d} pixelated or 3{\it d} voxelized cell geometries. 
\new{In this study, we only focus on the voxelized geometry of the isotropic voxel size in 3{\it d}.}
Further, the corresponding simulation kernel is easy to implement and maintain, and has low computational complexity with minimal problem of floating-point precision. On the other hand, the generated cell model has unrealistic surface-to-volume ratio due to its ``boxy" cell surface.

\begin{figure}[bt!]
\centering
	\includegraphics[width=0.4\textwidth]{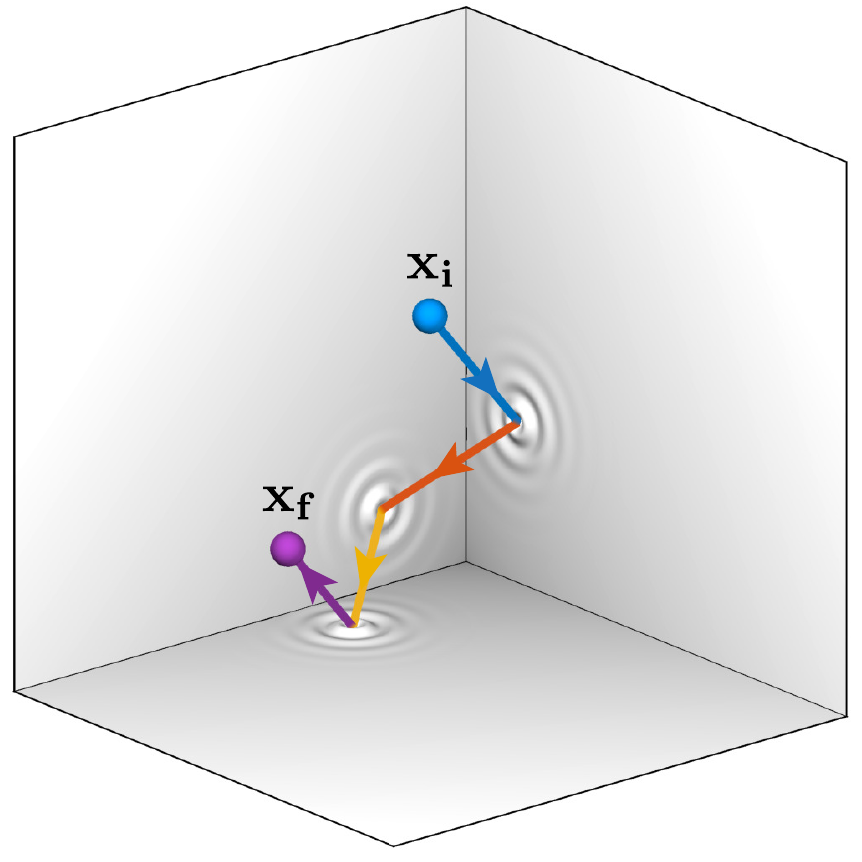}
	\caption{\textbf{In the voxelized geometry, a random walker has at most three interactions with membranes in each step.}
	The simulation of diffusion in 3{\it d} voxelized cell geometry can be simplified as {\it at most three} interactions between a random walker and membranes in each step, when the step size is smaller than the voxel size; the interactions include either an elastic collision with or a permeation through the membrane. This simplification is similar to the concept of the corner reflector (three consecutive coordinate reflections send vector ${\bf x} \to -{\bf x}$, after which no wall can be encountered within the corner, representing a given voxel), which has been applied to radar sets, glass prisms, and bicycle reflectors.}
	\label{fig:rms-elastic}
\end{figure}

To achieve fast and accurate MC simulations of diffusion in microscopy-based cell geometries, we chose to use the binary mask approach, where the random walker has {\it at most three}  interactions with membranes (elastic collision or membrane permeation) within each step, if the step size is smaller than the voxel size of the geometry (\figref{fig:rms-elastic}). \mpar{R3.2}\new{We define the voxel size as the length of the side of each 3$d$ cube.
Applying a small step size smaller than the voxel size within the voxelized geometry}, we only need to take the integer part of the walker's position (in a unit of voxel size) to determine where the walker resides after a particle-membrane interaction, minimizing the precision problem in numerical calculations. 

\mpar{R3.5}\new{Here we clarify that the six faces of each 3$d$ voxel do not necessarily represent the interface (i.e., cell membrane) of compartments; the membranes always coincide with faces in the voxelized geometry, but not all voxel faces are therefore part of membranes. When a random walker encounters a face between two voxels belonging to the same compartment, the random walker will permeate though the face as it does not exist; in contrast, when a random walker encounters a face between two voxels, each belonging to a different compartment, this face is effectively part of the cell membrane, and the check of permeation probability and particle-membrane interaction will be triggered (Sections~\ref{sec:interaction} and \ref{sec:membrane-permeation}).}

\mpar{R3.1}\new{A proper voxel size of the voxelized geometry should be (1) smaller than the length scale of cell shapes for accurate simulations, and (2) larger than the hopping step size in simulations to ensure at most three particle-membrane interactions in each step. For example, the voxel size of axon geometry should be much smaller than the axon diameter to capture the fine structures of cells, such as axon caliber variation and axonal undulation. However, choosing a small voxel size leads to an even smaller step size and a subsequently short time for each step (\secref{sec:membrane-permeation}), considerably increasing the number of steps and calculation time. To speed up simulations without losing the accuracy, it is recommended to choose the voxel size based on simulations in cell-mimicking simple geometries (e.g., cylinders for axons, spheres for cell bodies) with known analytical solutions of diffusion signals or metrics.}


Finally, for simulations of diffusion in a substrate consisting of multiple cells, the most computationally expensive calculation is to identify which compartment a random walker resides in. A way to solve this problem is to build a lookup table \citep{yeh2013table,fieremans2018cookbook}. \mpar{R3.3}\new{By partitioning cell geometries into many small voxels, the lookup table records compartment labels in each voxel. When a random walker hops across few voxels in a step, we only need to check compartments recorded in these voxels. Interestingly, each voxel in voxelized geometry records only one compartment label. In other words,} the voxelized geometry serves as the lookup table itself and thus saves the computational load and memory usage, which could be non-trivial for the GPU parallelization \citep{nguyen2018gpu}.

\subsubsection{Particle-membrane interaction: Why elastic collision is preferred to other approaches?}
\label{sec:interaction}

In MC simulations of diffusion, the diffusion process is discretized into multiple steps of constant length and random orientation in 3{\it d}, during a constant time duration for each step. To model the particle-membrane interaction, the most commonly used implementation for diffusion simulations is elastic collision \citep{szafer1995simulation,hall2009camino}, which properly equilibrates the homogeneous particle density around impermeable, permeable and absorbing membranes \citep{fieremans2018cookbook}. In RMS, we adopt this approach for simulations of high accuracy.

To reduce the computational load of simulations in complicated geometries, more and more studies apply other kinds of particle-membrane interactions. The first alternative is equal-step-length random leap (ERL) \citep{xing2013simulation}: a step encountering a membrane is canceled and another direction is chosen to leap until the new step does not encounter any membranes. In this way, ERL rejects some steps toward the membrane and effectively repels random walkers away from the membrane. This repulsive effect leads to inhomogeneous particle density around membranes (\ref{sec:app-erl}, \figref{fig:erl-bias-density}); therefore, ERL should not be applied to simulations of MR contrasts requiring homogeneous particle density, such as exchange, surface relaxation, and magnetization transfer. For the case of impermeable, non-absorbing membrane, the bias of diffusivity transverse and parallel to the membrane due to ERL is proportional to the step size (Figs.~\ref{fig:erl-bias-size}-\ref{fig:erl-bias-axial}) and could be controlled by choosing a small step size in simulations.

The second alternative for the interaction with membrane is rejection sampling \citep{ford1997simulation,waudby2011gpu,nguyen2018gpu,palombo2018leaflet}: a step encountering a membrane is canceled, and the random walker stays still for the step. This simple approach has been shown to be able to maintain a homogeneous particle density at all times \citep{szymczak2003rsampling} and is applicable to simulations of water exchange, as generalized in \ref{sec:app-rs}. However, it is rejecting some steps toward the membrane, which leads to a small bias in diffusivity parallel to membranes (\figref{fig:rs-bias-axial}), and thus it is still not the best option for an accurate simulation pipeline.

\mpar{R3.4}\new{To sum up, the criteria for choosing the particle-membrane interaction in an MC simulator include (1) the maintenance of homogeneous particle density around impermeable, permeable, and absorbing membranes, and (2) reliable simulations of diffusion metrics without interaction-related bias. Rejection sampling satisfies only the first criterion, and ERL fails both. In contrast, elastic collision satisfies both criteria, and thus we implement it in RMS for accurate simulations. Comparisons of other particle-membrane interactions can be found in \citep{szymczak2003rsampling,johannesson1996interaction}, where other approaches do not provide benefits in both calculation speed and accuracy.}

\subsubsection{Impermeable, permeable and absorbing membranes}
\label{sec:membrane-permeation}
When a random walker encounters an impermeable membrane within a step, the random walker is specularly reflected by the membrane, equivalently experiencing an elastic collision \citep{szafer1995simulation}. The displacement before and after the collision are summed up to the step size \citep{einstein1905diffusion}
\begin{equation} \label{eq:step-size}
    \delta s = \sqrt{2dD_0 \cdot \delta t}\,,
\end{equation}
where $d$ is dimensionality of space, $D_0$ is the intrinsic diffusivity, and $\delta t$ is the time of each MC step.

Further, when a random walker encounters a permeable membrane, the walker has a probability $1-P$ to be specularly reflected by the membrane, and a probability $P$ to permeate through. In \ref{sec:app-permeability}, we derive the connection between $P$ and the membrane permeability $\kappa$ in detail. Briefly, the permeation probability $P$ can be determined in two ways: 
\begin{enumerate}
    \item For the genuine membrane permeability $\kappa$, the permeation probability is given by
    \begin{subequations}
    \begin{align} \label{eq:perm-prob}
        P &\simeq \frac{\kappa_0 \delta s}{D_0}\cdot C_d\,,\\
        \new{C_d} &\equiv \begin{cases} 1 & d=1\,, \\ \pi/4 & d=2\,, \\ 2/3 & d=3\,, \end{cases}
    \end{align}
    \end{subequations}
    \mpar{R1.6}where $\kappa_0$ is the input permeability value 
    (with $\kappa_0\simeq \kappa$ if $P\ll 1$)
    \citep{powles1992simulation,szafer1995simulation,fieremans2018cookbook}. 
    When $P$ is not small, the relation of genuine permeability $\kappa$ and the input value $\kappa_0$ is given by \eqref{eq:kappa-correction}, leading to the probability of permeation from compartment 1 ($D_1$, $\delta s_1$) to compartment 2 ($D_2$, $\delta s_2$):
    \begin{equation} \label{eq:perm-prob-general}
        P_{1\to2} = \frac{\frac{\kappa \delta s_1}{D_1}\cdot C_d}{1+\frac{\kappa}{2}\left( \frac{\delta s_1}{D_1} + \frac{\delta s_2}{D_2}\right)\cdot C_d}\,.
    \end{equation}

    \item  Alternatively, without assigning a nominal permeability, the permeation probability can be determined by the ratio of tissue properties on both sides of the membrane, such as the intrinsic diffusivity and spin concentration, given by \eqref{eq:alt-perm-prob1-2} \citep{baxter2013simulation}. 
\end{enumerate}
The theoretical background and limitations of the two approaches are discussed in \ref{sec:app-permeability}.

Finally, for magnetization transfer (MT) effect, its MC simulation can be effectively modeled as an ``absorbing'' membrane, or as a surface relaxation effect. When a random walker encounters an absorbing membrane, the walker has a probability 
\begin{equation} \notag
    P_\text{MT}\simeq \frac{\varrho \, \delta s}{D_0}\cdot C_d
\end{equation}
to lose all its magnetization (to be saturated) \citep{fieremans2018cookbook},
where $C_d$ is the same constant as in \eqref{eq:perm-prob}, and $\varrho$ is the surface relaxivity (with the units of velocity). 
A random walker's magnetization is effectively ``saturated'' when the exchange with the macromolecule pool happens with the probability $P_\text{MT}$, 
equivalently introducing the weighting 
\begin{equation} \label{eq:wt-mt}
    \alpha_\varrho = \begin{cases} 1 & \text{not saturated;} 
    \\ 0 & \text{after saturation event} \end{cases}
\end{equation}
for the random walker's contribution to the net signal. 

The signal decay due to MT is caused by the exchange events between water protons and saturated macromolecule protons during diffusion. The mean-field estimate for the 
MT exchange rate $R$ (units of inverse time) between the liquid pool and the macromolecular pool, with exchange happening at the surfaces with the net surface-to-volume ratio $S/V$, is
given by \citep{slijkerman1998surface}
\begin{equation} \notag
    R = \varrho \cdot \frac{S}{V} \,.
\end{equation}

\subsubsection{Other MR contrast mechanisms}
\label{sec:contrast}
The transverse magnetization experiences the $T_2$ NMR relaxation. 
Hence, during all the time when the spin magnetization is in the transverse plane, such as for the conventional spin-echo diffusion sequence, the weighting due to the $T_2$ relaxation for each random walker's contribution to the overall signal is \citep{szafer1995simulation}
\begin{equation} \label{eq:wt-relax}
    \alpha_T = \exp\left(-\sum_i \frac{t_i}{T^{(i)}}\right)\,,
\end{equation}
where $t_i$ is the total time of staying in the $i$-th compartment during the echo time (TE), with the corresponding $T_2$ relaxation time $T^{(i)} \equiv T_2^{(i)}$. Similarly, for simulations of a stimulated-echo sequence, the  weighting due to the $T_1$ relaxation during the mixing time (when the magnetization is parallel to $B_0$) for each random walker's contribution is also given by \eqref{eq:wt-relax} \citep{woessner1961ste}, where $t_i$ is the total time of staying in the $i$-th compartment during the mixing time, with the corresponding $T_1$ relaxation time $T^{(i)}\equiv T_1^{(i)}$.

Other contrasts, such as susceptibility effect, blood oxygen level dependent (BOLD) effect and  intravoxel incoherent motion (IVIM), can be further added to RMS by modifying each random walker's phase factor $e^{-i\varphi(t)}$ due to the 
flow velocity and the local Larmor frequency offset 
in a standard way. 


\subsubsection{Parallel computations and input-output}
\label{sec:gpu}
To achieve precise simulations of diffusion in complicated 3{\it d} microgeometry, it is inevitable to employ a large number of random walkers. However, the complexity of substrate and the combination of multiple MR contrast mechanisms in recent studies substantially increase the computational load in simulations, prompting the usage of parallel computations, either through multiple nodes on a cluster, multiple threads on a CPU/GPU core, or their combinations \citep{waudby2011gpu,xu2018susceptibility,palombo2018leaflet,nguyen2018gpu,lee2020axial}.

The simulation in RMS is implemented in CUDA C++ and accelerated through the parallel computation on GPU. Furthermore, by performing multiple simulations on a GPU cluster, it is possible to further accelerate simulations through multiple nodes, with each node equipped with a GPU core.

The input RMS files and the shell script to compile and run the CUDA C++ kernel are all generated in a MATLAB script. Moreover, both the input and output files are text files, and all the codes are open-sourced and platform-independent. These properties make the RMS congenial to even beginners in this field to adapt RMS and program their own simulation pipelines.
\mpar{R1.3}\new{In \tabref{tab:input}, we summarize input parameters of RMS.}

\subsection{Monte Carlo simulations in realistic microstructure}
\label{sec:example}

Here we shortly summarize the diffusion simulations implemented in RMS:
\begin{enumerate}
    \item Random walkers' initial positions ${\bf x}_0$ are randomly initialized to achieve a homogeneous particle density. In RMS, we provide users the flexibility to define a ``dead'' space, where no random walkers are initialized and allowed to step in.
    
    \item Random walkers diffuse in a continuous space with voxelized microgeometry (\secref{sec:substrates}). For the $i$-th step, the random walker is at position ${\bf x}_{i-1}$ before the random hop, and a step vector $\new{\delta}{\bf x}$ of constant length $\delta s$ ($<$ voxel size) and random direction in 3{\it d} is generated. \mpar{R1.4}
    
    \item The random walker hops to a new position 
    \begin{equation} \label{x_i}
        {\bf x}_i = {\bf x}_{i-1}+\new{\delta} {\bf x}
    \end{equation}
   if it does not encounter the edge of the voxel where it resides in before the hopping. 
   If the edge of the voxel is encountered, it will either permeate though or be elastically reflected from the voxel edge based on the permeation probability defined in \secref{sec:interaction}, leading to a new position ${\bf x}_i$ accordingly. The permeation probability is set to 1 if the random walker encounters the edge between two adjacent voxels belonging to the same compartment.
    
    \item The above particle-membrane interaction will repeat at most three times during each MC step due to the voxelized geometry and a small step size ($<$ voxel size), as illustrated in \figref{fig:rms-elastic}. 
    
    \item The diffusion metrics and signals are then calculated based on the diffusional phase due to simulated diffusion trajectory ${\bf x}_i$ and signal weighting ($\alpha_\varrho$, $\alpha_T$ in Sections~\ref{sec:membrane-permeation} and \ref{sec:contrast}), with details in the following \secref{sec:metrics}.
\end{enumerate}

The accuracy of RMS is guaranteed by the theoretical exploration of particle-membrane interaction (\ref{sec:app-erl} and \ref{sec:app-rs}) and water exchange (\ref{sec:app-permeability}).

\subsection{Diffusion metrics}
\label{sec:metrics}
Normalized diffusion signals of the pulse-gradient spin echo sequence are calculated based on the diffusional phase $\varphi$ accumulated along the diffusion trajectory ${\bf x}(\tau)$ \citep{szafer1995simulation}:
\begin{equation} \label{eq:signal}
    S = \frac{\langle \alpha\cdot e^{-i\varphi}\rangle_N}{\langle \alpha \rangle_N}\,,\quad \varphi = \int_0^\text{TE} {\bf g}(\tau)\cdot {\bf x}(\tau) \, \text{d} \tau\,,
\end{equation}
where $\langle...\rangle_N$ is the average over $N$ random walkers, $\alpha=\alpha_\varrho\cdot \alpha_T$ is the signal weighting of individual walker in Eqs.~(\ref{eq:wt-mt}) and (\ref{eq:wt-relax}), 
$\langle \alpha\rangle_N$ is the normalization constant (the total non-diffusion-weighted signal), ensuring that the signal $S\equiv 1$ in the absence of diffusion weighting $(b=0)$, 
TE is the echo time, and ${\bf g}(\tau)$ is the diffusion-sensitizing gradient of the Larmor frequency.

For the pulsed-gradient sequence, the cumulant expansion of \eqref{eq:signal} yields the diffusivity and kurtosis in the \textit{narrow pulse} limit, given by
\begin{equation*}
    D = \frac{1}{2t}\cdot\frac{ \left\langle \alpha\cdot\left(\Delta {\bf x} \cdot \hat{\bf g}\right)^2 \right\rangle_N}{\langle \alpha \rangle_N}\,,\quad
    K = \frac{ \left\langle \alpha\cdot\left(\Delta {\bf x} \cdot \hat{\bf g}\right)^4 \right\rangle_N }{ \left\langle \alpha\cdot\left(\Delta {\bf x} \cdot \hat{\bf g}\right)^2 \right\rangle_N ^2} - 3\,,
\end{equation*}
where $\Delta {\bf x}={\bf x}(t)-{\bf x}(0)$ is the diffusion displacement, $\hat{\bf g}$ is the gradient direction, and $t$ is the diffusion time. For an ideal pulse-gradient sequence, $\text{TE}=t$ in the narrow pulse limit.

For the pulsed-gradient sequence with \textit{wide pulses}  \citep{stejskal1965pgse}, to obtain the diffusion and kurtosis tensors we simulate signals for multiple diffusion weightings $b$ according to the gradient wave form ${\bf g}(\tau)$, and fit the cumulant expansion in the powers of $b$ to the simulated signals \citep{jensen2005dki}:
\begin{multline} \notag
    \ln S \simeq -b \sum_{ij}n_{ij} D_{ij}\\
    + \tfrac{1}{6}(b \bar{D})^2 \sum_{ijkl}n_{ijkl} W_{ijkl} + {\cal O}(b^3)\,,\quad b=g^2\delta^2(\Delta -\delta/3)\,,
\end{multline}
where $n_{ij}=\hat{g}_i\hat{g}_j$, $n_{ijkl}=\hat{g}_i\hat{g}_j\hat{g}_k\hat{g}_l$, $D_{ij}$ and $W_{ijkl}$ are diffusion and kurtosis tensors, $\bar{D}=\frac{1}{3}\sum_i D_{ii}$, 
and the diffusion time $t\approx \Delta$ is roughly the time interval $\Delta$ between two gradient pulses with the precision of its definition given by  the gradient pulse width $\delta$ \citep[Sec.~2]{novikov2019review}.
The above formula and the overall diffusion attenuation $S=S[{\bf g}(\tau)]$ can be generalized to be a {\it functional} of the multi-dimensional gradient trajectory ${\bf g}(\tau)$, with $b n_{ij} \to B_{ij}$ being the $B$-matrix, and with the higher-order terms defiend accordingly \citep{Topgaard2017}. 

\begin{table*}[t!]
\centering
\mnote{R1.3}
\begin{tabular}{ lll } 
 \hline
 Voxel size & $l_p$ & $<$ length scale in cell geometry (e.g., axon diameter, cell body size) \\
 Intrinsic diffusivity & $D_0$ & $\sim 2-3$ $\mu$m$^2$/ms in vivo  \\
 Time for each step & $\delta t$ & small enough such that $P_{1\to2}\ll1$, $P_\text{MT}\ll1$, and $\delta s<l_p$\\
 Total time & $T_\text{max}$ & $>t$ for narrow PG, $>(t+\delta)$ for wide PG \\
 Number of steps & $N_\text{step}$ & $=\left\lceil T_\text{max}/\delta t \right \rceil$, at least 1000 steps to lower discretization error\\
 Number of particles & $N_\text{par}$ & as many as possible within the given calculation time \\
 Step size & $\delta s$ & $=\sqrt{2dD_0 \cdot \delta t}$, $d=$dimension (typically $d=3$) \\
 \hline \hline
 \textbf{Optional} && \\
 Diffusion gradient & ${\bf g}(\tau)$ & limited by maximal gradient strength, slew rate, and b-value \\
 Permeability & $\kappa$ & \\
 Surface relaxivity & $\varrho$ & \\
 MT exchange rate & $R$ & $=\varrho\cdot S/V$ \\
 Relaxation time & $T_1$/$T_2$ & $\sim 800/80$ ms in brain white matter at 3T\\
 \hline
\end{tabular}
\caption{\textbf{Input parameters of RMS.} The guideline of setting up simulation parameters is based on \citep{fieremans2018cookbook}. MT: magnetization transfer, PG: pulsed-gradient, $t$: diffusion time, $\delta$: gradient pulse width, $S/V$: surface-to-volume ratio.}
\label{tab:input}
\end{table*}

\section{RMS Applied to Intra-Axonal Microstructure}
\label{sec:results}

In this Section, we describe an RMS-compatible example of a realistic electron microscopy (EM) tissue segmentation (\secref{sec:segmentation}), 
give an overview of the related biophysical models (\secref{sec:models}), 
describe the RMS settings for MC in axonal geometry (\secref{sec:MC-IAS}), and outline our results for the diffusion along (\secref{sec:para}) and transvere (\secref{sec:perp}) to the axons.

All procedures performed in studies involving animals were in accordance with the ethical standards of New York University School of Medicine. All mice were treated in strict accordance with guidelines outlined in the National Institutes of Health Guide for the Care and Use of Laboratory Animals, and the experimental procedures were performed in accordance with the Institutional Animal Care and Use Committee at the New York University School of Medicine. This article does not contain any studies with human participants performed by any of the authors.

\subsection{Axon segmentation based on electron microscopy}
\label{sec:segmentation}

A female 8-week-old C57BL/6 mouse was perfused trans-cardiacally, and the genu of corpus callosum was fixed and analyzed with a scanning electron microscopy (Zeiss Gemini 300). We selected a subset of the EM data ($36\times48\times20$ $\mu$m$^3$ \mpar{R1.8}\new{in volume, $6\times6\times100$ nm$^3$ in voxel size}), down-sampled its voxel size to \new{$24\times24\times100$ nm$^3$ slice-by-slice by using Lanczos interpolation}, and segmented the intra-axonal space (IAS) of 227 myelinated axons using a simplified seeded-region-growing algorithm \citep{adams1994srg}. \new{The segmented IAS mask was further down-sampled into an isotropic voxel size of ($100$ nm)$^3$ for numerical simulations.} More details can be found in our previous work \citep{lee2019em}. For simulations in RMS, the voxelized microgeometry based on the axon segmentation is shown in the bottom-right corner of \figref{fig:history} \citep{lee2020axial}. In the following examples, we assume that the tissue properties (i.e., diffusivity, relaxation time) are the same in cytoplasm and organelles (e.g., mitochondria).

\subsection{Tissue parameters and biophysical models for the axonal geometry}
\label{sec:models}

To quantify the axon geometry, we define axon's inner diameter, caliber variation, and axonal undulation as follows: The inner diameter $2r \equiv2\sqrt{\Omega/\pi}$ of an axon cross-section is defined as that of an equivalent circle with the same cross-sectional area $\Omega$ \citep{west2016gratio}; the caliber variation is defined as the coefficient of variation of radius (ratio of standard deviation to mean), $\text{CV}(r)$ \citep{lee2019em}; the axonal undulation $|{\bf w}|$ is defined as the shortest distance between the axonal skeleton and the axon's main axis \citep{lee2020radial}; based on the analysis of an axonal skeleton, we can roughly estimate the length scale of undulation amplitude $w_0$ and wavelength $\lambda$ using a simplified 1-harmonic model (\ref{sec:app-undulation}).

In our previous studies, we showed that restrictions {\it along axons} are randomly positioned with a finite correlation length $\ell_c$ of a few micrometers \citep{lee2020axial}. This short-range disorder leads to the diffusivity along axons 
\begin{equation} \label{eq:D-axial}
    D_\parallel(t) \simeq D_\infty + c\cdot t^{-1/2} \,, 
    \quad t\gg t_c= {\ell_c^2 \over 2D_\infty}
\end{equation}
approaching its $t\to\infty$ limit $D_\infty$ in a power-law fashion with exponent $1/2$ \citep{novikov2014meso,fieremans2016invivo,lee2020axial}. 
Here, $c$ is the strength of the restrictions; the relation (\ref{eq:D-axial}) holds in the narrow pulse limit, and acquires further corrections for finite pulse width $\delta$. The scaling (\ref{eq:D-axial}) becomes valid when the diffusion length exceeds the correlation length $\ell_c$ of the structural disorder (e.g., beads) along axons.

The bulk diffusivity along axons correlates with axon geometry via \citep{lee2020axial}
\begin{equation} \label{eq:zeta-cv}
    \zeta \equiv \frac{D_0-D_\infty}{D_\infty} \propto \text{CV}^2(r)\,,
\end{equation}
indicating that the stronger the caliber variation $\text{CV}(r)$, the smaller $D_\infty$.

Furthermore, the nonzero diffusivity {\it transverse to axons} is contributed by axon caliber and undulation  \citep{nilsson2012undulation,brabec2020undulation,lee2020radial}. In the wide pulse limit, untangling the two effects in diffusion transverse to axons is non-trivial. Instead, we translate the radial diffusivity $D_\perp$ into the effective radius measured by MR, $r_\text{MR}$ \citep{burcaw2015meso,veraart2020highb}, and identify which contribution dominates \citep{lee2020radial}:
\begin{align} \label{eq:r-mr}
    r_\text{MR} &\equiv \left[\frac{48}{7}\delta\left(t-\frac{\delta}{3}\right)D_0D_\perp\right]^{1/4} \\ \label{eq:r-cal-und}
    &\simeq  \begin{cases} r_\text{cal}\equiv \left(\langle r^6\rangle/\langle r^2\rangle\right)^{1/4}\,, & t_D\ll\delta\ll t_u \\ 
    r_\text{und}\equiv \left(\frac{6}{7\pi^2}\right)^{1/4} \cdot \sqrt{w_0\lambda}\,, & \delta \gg t_u \end{cases}
\end{align}
where $r_\text{cal}$ and $r_\text{und}$ are the apparent axon size due to the axon caliber and undulation respectively, and
\begin{equation} \label{tc}
t_D\equiv r^2/D_0 \quad \mbox{and}\quad t_u\equiv \lambda^2/(4\pi^2D_0) 
\end{equation}
are correlation times for the axon caliber and undulation respectively. In realistic axonal shapes, $r\sim 0.5$ $\mu$m and $\lambda\sim$ 20 $\mu$m lead to $t_D\sim0.1$ ms and $t_u\sim5$ ms for $D_0=2$ $\mu$m$^2$/ms \citep{lee2019em,lee2020radial}.

To  factor out the effect of undulation, diffusion signals are spherically (orientationally) averaged for each $b$-shell. Assuming that axonal segments have cylindrical shapes, the spherically averaged intra-axonal signals are given by \citep{jensen2016fiberball,veraart2019highb,veraart2020highb}
\begin{equation} \label{eq:S-sph}
    \bar{S}(b) \simeq \beta\, e^{-b D_\perp + {\cal O}(b^2)}\,b^{-1/2} \,,
\end{equation}
where $\beta\equiv\sqrt{\pi/(4D_a)}$, and $D_a$ is the diffusivity along axon segments within the diffusion length $\sim\sqrt{D_a t}$. The effective radius is then calculated based on $D_\perp$ and \eqref{eq:r-mr}.

\subsection{RMS settings for MC simulations in the intra-axonal space}
\label{sec:MC-IAS}

To demonstrate the relation of diffusion metrics and microgeometry in intra-axonal space in Eqs.~(\ref{eq:zeta-cv}) and (\ref{eq:r-cal-und}), we performed diffusion simulations in 227 realistic axons segmented from mouse brain EM in \secref{sec:segmentation}: random walkers diffused over $1\times10^6$ steps with a duration $\delta t=2\times10^{-4}$ ms and a length $\delta s=0.049$ $\mu$m (\eqref{eq:step-size} with $D_0=2$ $\mu$m$^2$/ms) for each of the following simulations of monopolar pulsed-gradient sequences.

For simulations of narrow pulse sequence, $1\times10^7$ random walkers per axon were applied. The apparent diffusivity $D_\parallel(t)$ along axons was calculated by using diffusion displacement along the axons' main direction (z-axis) at diffusion time $t\leq 100$ ms.  \eqref{eq:D-axial} for the time range $t=60-80$ ms was fit to the simulated $D_\parallel(t)$, and the fit parameter $D_\infty$ was correlated with the caliber variation $\text{CV}(r)$ in \eqref{eq:zeta-cv}.

For simulations of wide pulse sequence, $2\times10^4$ random walkers per axon were applied. The duration between pulses $\Delta=1-100$ ms was equal to the gradient pulse width $\delta$, such that diffusion time $t\approx \Delta=\delta$. 
The diffusion signals were calculated based on the accumulated diffusion phase (\ref{eq:signal}) for 10 b-values $=0.2-2$ ms/$\mu$m$^2$ along directions (x- and y-axes) transverse to axons' main direction (z-axis). The simulated apparent diffusivity $D_\perp$ transverse to each axon was translated to the effective radius by using \eqref{eq:r-mr} and compared with the contribution of caliber variations and axonal undulations in \eqref{eq:r-cal-und}.

For simulations of directionally averaged signal, $2\times10^4$ random walkers per axon were applied. The diffusion time and gradient pulse width $(t,\delta)=(20,7.1)$, $(30,13)$ and $(50.9,35.1)$ ms were used to match the experimental settings on animal 16.4T MR scanner (Bruker BioSpin), Siemens Connectome 3T MR scanner, and clinical 3T MR scanner (Siemens Prisma) \citep{veraart2020highb,lee2020radial}. Diffusion signals were calculated based on the accumulated diffusional phase for 18 b-values $=16-100$ ms/$\mu$m$^2$ along 30 uniformly distributed directions for each b-shell. \eqref{eq:S-sph} was fit to the spherically averaged signal $\bar{S}(b)$ from all axons (volume-weighted sum), 
and the fit parameter $D_\perp$ was again translated to the effective radius in \eqref{eq:r-mr} and compared with the contribution of caliber variations and axonal undulations of all axons, i.e., $r_\text{cal}$ in \eqref{eq:r-cal-und} and $\tilde{r}_\text{und}$ defined below. 

The $r_\text{und}$ in \eqref{eq:r-cal-und} is defined for the individual axon and does not take the volume differences between axons into account. To add up the contribution of $D_\perp$ due to undulations of all axons, we define a volume-weighted average of $r_\text{und}$:
\begin{equation} \label{eq:r-und-vw}
    \tilde{r}_\text{und} = \left(\sum_i f_i \cdot r_{\text{und},i}^4\right)^{1/4}\,,
\end{equation}
with the $i$-th axon's volume fraction $f_i$, such that $\sum_i f_i = 1$.

\begin{figure}[bt!]
\centering
	\includegraphics[width=0.45\textwidth]{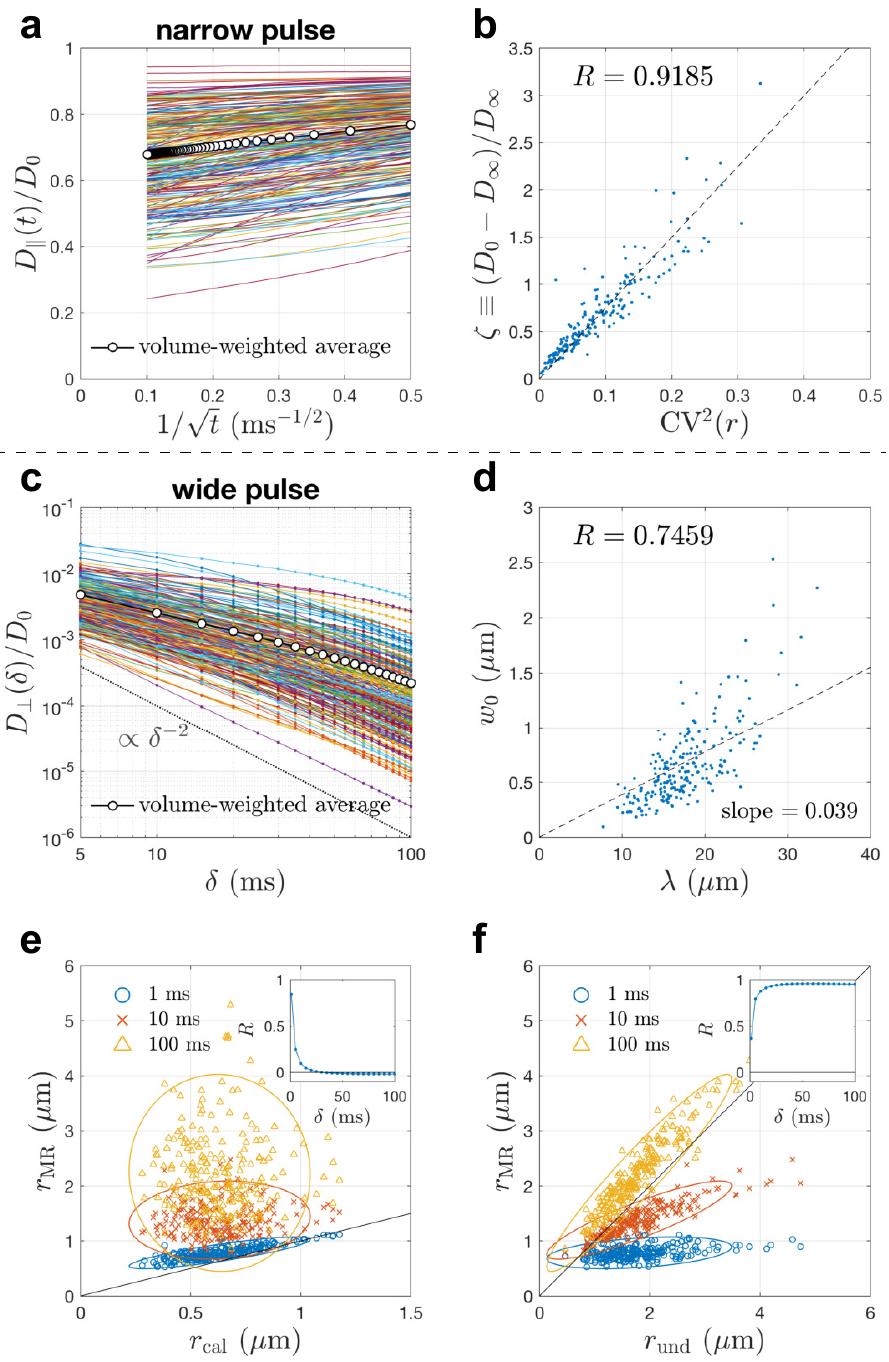}
	\caption{\textbf{Simulated apparent diffusivity parallel and transverse to 227 realistic axons segmented from a mouse brain corpus callosum.}
	\textbf{a-b} In the narrow pulse limit, the simulated time($t$)-dependent diffusivity $D_\parallel(t)$ along axons scales as $1/\sqrt{t}$ in \eqref{eq:D-axial}, and its bulk diffusivity $D_\infty$ correlates with the caliber variation $\text{CV}(r)$ via \eqref{eq:zeta-cv}. \textbf{c} In the wide pulse limit ($t=\delta$), the simulated diffusivity $D_\perp(\delta)$ transverse to axons scales as $\delta^{-2}$ (dashed line) at very long times $\delta \gtrsim 50$ ms. \textbf{d} Based on the analysis of realistic axonal skeleton (\ref{sec:app-undulation}), the estimated undulation wavelength $\lambda$ positively correlates with the undulation amplitude $w_0$. \textbf{e-f} Translating the simulated $D_\perp$ into an effective radius $r_\text{MR}$ measured by MR via \eqref{eq:r-mr}, the value of $r_\text{MR}$ is compared with the contributions of caliber variations and undulations, $r_\text{cal}$ and $r_\text{und}$ in \eqref{eq:r-cal-und} respectively. The Pearson's correlation coefficient $R$ of $r_\text{MR}$ and each contribution is also shown as a function of time $\delta$ (insets). The result in panel \textbf{b} is similar to Fig. 2d in \citep{lee2020axial}, whereas the short $T_2$ and slow diffusivity in mitochondria are ignored here to focus on the effect of axonal shape. The results in panels \textbf{c} and \textbf{e-f} are similar, but not identical, to the ones obtained using ERL algorithm in Fig. 7 of \citep{lee2020radial}. Here, these simulations were re-done using elastic collision with impermeable membranes, and the estimation of undulation amplitude and wavelength is purely based on the geometry of axonal skeleton, as opposed to using ERL in simulations and fitting a 1-harmonic undulation model to simulated $D_\perp$ with fit parameters $w_0$ and $\lambda$ in \citep{lee2020radial}. The effective theory parameters and scaling relations are essentially the same for both MC implementations (elastic collision vs ERL) because only diffusion simulations within impermeable cells were performed.
	\rem{complicated model-- uninformative, describe a model in 1-2 words? If no space, move all this description of differences bw Lee2020 to main text. 1. The 1-harmonic undulation model is added.}
	}
	\label{fig:ad-rd}
\end{figure}

\subsection{Results: Diffusion metrics and caliber variations along axons}
\label{sec:para}

In 227 realistic axons segmented from a mouse brain, the simulation result of narrow pulse sequence shows that the apparent diffusivity $D_\parallel(t)$ along axons scales as $1/\sqrt{t}$ in \eqref{eq:D-axial} (\figref{fig:ad-rd}a),
and the fitted bulk diffusivity $D_\infty$ correlates with the caliber variation $\text{CV}(r)$ in \eqref{eq:zeta-cv} (\figref{fig:ad-rd}b). This demonstrates that the diffusion along axons is characterized by the short-range disorder in 1{\it d}, corresponding to randomly positioned beads along axons \citep{novikov2014meso,fieremans2016invivo,lee2020axial}.
\rem{what happens if average over all axons to better see $t^{-1/2}$? 1. The average is shown in Fig. 3a and 3c now.}


\subsection{Results: Diffusion metrics transverse to axons}
\label{sec:perp}

\begin{figure}[bt!]
\centering
	\includegraphics[width=0.42\textwidth]{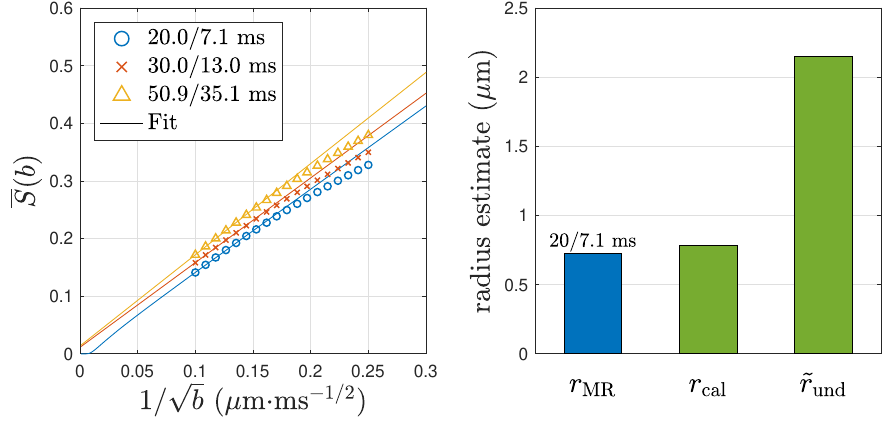}
	\caption{\textbf{Simulated spherically averaged signal of all 227 realistic axons segmented from a mouse brain corpus callosum.}
	The simulated spherically averaged signal $\bar{S}(b)$ of all axons scales as $1/\sqrt{b}$ in \eqref{eq:S-sph}, and its negative intercept at $b\to\infty$ indicates an estimate of diffusivity $D_\perp$ transverse to axons. For $(t,\delta)=(20, 7.1)$ ms (animal scanner), the negative intercept of $\bar{S}(b)$ corresponds to a $D_\perp\simeq 1.6\times10^{-4}$ $\mu$m$^2$/ms and an MR-measured radius $r_\text{MR}\simeq0.72$ $\mu$m based on \eqref{eq:r-mr}; however, for $(t,\delta)=(30,13)$ and $(50.9,35.1)$ ms (human scanner), the positive intercepts of $\bar{S}(b)$ leads to the non-physical, negative $D_\perp$. The MR-measured radius $r_\text{MR}$ at $(20,7.1)$ ms is close to the caliber contribution from histology, $r_\text{cal}=0.79$ $\mu$m in \eqref{eq:r-cal-und}, and much smaller than the undulation contribution from histology, $\tilde{r}_\text{und}=2.15$ $\mu$m in \eqref{eq:r-und-vw}, indicating that the undulation contribution is factored out by orientational average. The result of left panel is similar, but not identical, to the ones obtained using ERL algorithm in Fig. 8g of \citep{lee2020radial}. Here, these simulations were re-done using elastic collision with impermeable membranes. The effective theory parameters and scaling relations are essentially the same for both MC implementations because only simulations within impermeable axons were performed.
	}
	\label{fig:sig-sph}
\end{figure}

We compare the contribution of the axon caliber and axonal undulation to the diffusivity $D_\perp$ transverse to axons \citep{lee2020radial}: At clinical diffusion times ($t\sim\delta>10$ ms) of conventional wide-pulse sequences, the apparent diffusivity transverse to axons is dominated by the contribution of undulations (\figref{fig:ad-rd}c-f). To release the requirement of very short diffusion time $\sim 1$ ms for accurate axon diameter mapping, the modeling of spherically averaged signals at multiple diffusion weightings are employed to partially factor out the undulation effect (\figref{fig:sig-sph}).

\subsubsection{Diffusivity and axonal undulations}
In realistic axons of a mouse brain, the simulation result of wide pulse sequences shows that the apparent diffusivity $D_\perp$ transverse to axons scales as $\delta^{-2}$ at very long time $\delta \gtrsim 50$ ms in most axons (\figref{fig:ad-rd}c). 
This scaling is consistent with both the undulation and caliber contributions, since 
\[
D_\perp \simeq -\tfrac{1}{b}\ln S 
\propto {r_{\rm cal}^4 \delta/D_0 \over  \delta^2 (\Delta-\delta/3)}
\sim \delta^{-2}\,, \quad \Delta=\delta \,.
\]
However, the {\it onset} of this scaling depends on the correlation time, \eqref{tc}. 
If the caliber dominates, the $\delta^{-2}$ scaling will occur 
for $\delta \gg t_D\sim 0.1$ ms \citep{neuman1974pore}.
However, the actual scaling happens at times about 2 orders of magnitude longer. 
The  onset for the undulations, $\delta\gg t_u  \sim 5$ ms, as estimated after \eqref{tc}, is far more consistent with our RMS results.  

To better understand the length scale of undulations, the undulation amplitude $w_0$ and wavelength $\lambda$ are estimated based on the axonal skeleton (\ref{sec:app-undulation}) (\figref{fig:ad-rd}d). The scatter plot of $w_0$ and $\lambda$ shows positive correlation; in other words, the longer undulation wavelength, the stronger undulation amplitude. Furthermore, the slope $\text{d}w_0/\text{d}\lambda\simeq 0.039$ and the simplified 1-harmonic undulation model in \eqref{lambda-w0} indicate an estimate of the \textit{intra-axon} undulation dispersion $\simeq 10^\circ$, which is about a factor of 2 smaller than the \textit{inter-axon} fiber orientation dispersion $\simeq 22^\circ$, leading to an overall dispersion angle $\simeq \sqrt{(10^{\circ})^2 + (22^{\circ})^2} \simeq 24^\circ$ (\ref{sec:app-undulation}) \citep{ronen2014cc,lee2019em}. Interestingly, the inter-axon dispersion $\sim22^\circ$ and overall dispersion $\sim24^\circ$ are consistent with the numerical calculation of time-dependent dispersion at long and short times respectively in \citep{lee2019em}, where the same group of axons were analyzed.
\rem{see if you agree with par added in App D; maybe formulate the result about undul contribution to dispersion here? 1. Calculation and discussion are added.}

Furthermore, the simulated $D_\perp$ is translated to the effective radius $r_\text{MR}$ measured by MR via \eqref{eq:r-mr} and compared with the contribution of axon caliber and axonal undulation, i.e., $r_\text{cal}$ and $r_\text{und}$ in \eqref{eq:r-cal-und} respectively (\figref{fig:ad-rd}e-f). At very short time $\delta \sim 1$ ms, the MR estimate $r_\text{MR}$ coincides with the contribution of axon caliber $r_\text{cal}$; however, at long time $\delta\sim 100$ ms, the MR estimate $r_\text{MR}$ is consistent with the contribution of axonal undulations $r_\text{und}$. Finally, at clinical time range $\delta \sim 10-50$ ms, the contribution of undulations $r_\text{und}$ shows much higher correlation with the MR estimate $r_\text{MR}$, compared with the correlation of $r_\text{MR}$ and $r_\text{cal}$ (the insets of \figref{fig:ad-rd}e-f). These observations demonstrate that axonal undulations confound the axonal diameter mapping at clinical time range and need to be factored out for an accurate diameter estimation.

\subsubsection{Spherically averaged signals and axonal diameter mapping}
The spherically averaged signal $\bar{S}(b)$ of all axons from a mouse brain scales as $1/\sqrt{b}$ in \eqref{eq:S-sph} (\figref{fig:sig-sph}), and its negative intercept at $b\to\infty$ corresponds to an estimate of radial diffusivity $D_\perp$, whose contribution of undulations is partially factored out: On the one hand, at short time $(t,\delta)=(20, 7.1)$ ms (animal scanner), the $\bar{S}(b)$ has a negative intercept, indicating a $D_\perp\simeq 1.6\times 10^{-4}$ $\mu$m$^2$/ms and an MR-measured radius $r_\text{MR}\simeq 0.72$ $\mu$m in \eqref{eq:r-mr}, which is close to the histology-based caliber contribution $r_\text{cal}=0.79$ $\mu$m in \eqref{eq:r-cal-und} and smaller than the histology-based undulation contribution $r_\text{und}=2.15$ $\mu$m in \eqref{eq:r-und-vw}. 
The fitted diffusivity along axon segments, $D_a\simeq 0.19 \cdot D_0$, is unexpectedly low since the variation of local diffusivities along axons is not considered in \eqref{eq:S-sph}.
On the other hand, at longer diffusion times $(t,\delta)=(30,13)$ ms and $(50.9, 35.1)$ ms (human scanners), the $\bar{S}(b)$ has a positive intercept, leading to the non-physical, negative $D_\perp$.

The simulation result demonstrates that the requirement of short diffusion time $\sim1$ ms for an accurate axon size estimation can be partly released by spherically averaging signals.

\section{Outlook}
\label{sec:discussion}
Performing MC simulations in realistic cell geometries using the proposed RMS helps to test the sensitivity of diffusion MRI to tissue features and validate the biophysical models. RMS is an open-source platform for the Monte Carlo simulations of diffusion in realistic tissue microstructure. 
\mpar{R1.9}\new{In addition to the examples of diffusion within intra-cellular space, it is also possible to perform simulations of diffusion in the extra-cellular space, as well as of the exchange between intra- and extra-cellular spaces. The tissue preparation preserving extra-cellular space in histology is non-trivial, prompting the development of pipelines to generate semi-realistic tissue microstructure by packing multiple artificial ``cells", such as MEDUSA \citep{ginsburger2019medusa} and ConFiG \citep{callaghan2020config}, whose generated microgeometry potentially could be transformed to 3{\it d} voxelized data, compatible with RMS.}

Beyond the proposed pipeline, here we provide an outlook for the next generation simulation tools.
So far, most of the implementations of diffusion simulations focus on the MR sequence with radiofrequency pulses of 90$^\circ$ and 180$^\circ$, such as the spin-echo \citep{stejskal1965pgse} and stimulated-echo \citep{tanner1970pgste} sequences. To simulate  other MR sequences, such as the steady-state free precession sequence \citep{mcnab2008ssfp}, it is required to combine the diffusion simulation with the Bloch simulator in MR system. Furthermore, the generation of accurate cell segmentation for simulations in microscopy-based geometry is time-consuming and labor-intensive; performing diffusion simulations directly in a mesoscopic diffusivity map, obtained via a transformation of the microscopy intensity, could largely simplify the validation pipeline --- however, such segmentation-less simulation must itself be thoroughly validated.

As tissue microstructure imaging with MRI is a vastly expanding research area of quantitative MRI, the synergies between microstructure model validation, hardware and acquisition techniques have a potential to transform MRI into a truly quantitative non-invasive microscopy/histology technique, as discussed in \citep{novikov2020future}. Hence, the development of next-generation simulation tools will benefit not only the field of microstructural MR imaging, but  the whole MRI community.

\section{Conclusions}
\label{sec:conclusion}
Numerical simulations in realistic 3\textit{d} microgeometry based on microscopy data serve as a critical validation step for biophysical models, in order to obtain quantitative biomarkers, e.g., axonal diameter, the degree of caliber variations and axonal undulations, for potential clinical applications. With the help of the proposed RMS pipeline, it is possible to achieve fast and accurate simulations of diffusion in realistic tissue microstructure, as well as the interplay with other MR contrasts. RMS enables {\it ab initio} simulations from realistic microscopy data, facilitating  model validation and experiment optimization for microstructure MRI in both clinical and preclinical settings.

\section*{Acknowledgements}

We would like to thank Sune Jespersen for  fruitful discussions about  theory, and the BigPurple High Performance Computing Center of New York University Langone Health for numerical computations on the cluster. Research was supported by the National Institute of Neurological Disorders and Stroke of the NIH under awards R01 NS088040 and R21 NS081230, by the National Institute of Biomedical Imaging and Bioengineering (NIBIB) of the NIH under award number U01 EB026996, and by the Irma T. Hirschl fund, and was performed at the Center of Advanced Imaging Innovation and Research (CAI2R, www.cai2r.net), a Biomedical Technology Resource Center supported by NIBIB with the award P41 EB017183.

\figref{fig:history} is adapted from \citep{chin2002simulation,xu2018susceptibility,nguyen2018gpu,palombo2019astrocyte,lee2020axial} with permission from Wiley, Elsevier, and Springer Nature.

\figref{fig:solid-angle} is adapted from \citep{fieremans2018cookbook} with permission from Elsevier.

\section*{Declarations of interest}
None.

\section*{Data and code availability statement}
The SEM data and axon segmentation can be downloaded on our web page (www.cai2r.net/resources/software). 

\mpar{R1.7}
The source codes of Monte Carlo simulations can be downloaded on our Github page \new{(github.com/NYU-DiffusionMRI/monte-carlo-simulation-3D-RMS)}. The first release of RMS supports simulations of multiple MR contrasts (diffusion, $T_2$ relaxation, water exchange) and the pulsed-gradient spin-echo sequence. Simulations of other MR contrasts (MT and $T_1$ relaxation), sequences (stimulated-echo), arbitrary diffusion gradient waveforms, as well as flow and susceptibility-induced effects will be updated in the future.



\appendix
\section{Unbiased simulation of particle-membrane interaction --- equal-step-length random leap}
\renewcommand{\theequation}{A.\arabic{equation}}
\setcounter{equation}{0}
\setcounter{figure}{0}
\label{sec:app-erl}

To largely reduce the computational load of MC simulations, \citet{xing2013simulation} proposed to model the interaction of random walkers and membranes by using the equal-step-length random leap (ERL): a step crossing the membrane is canceled, and another direction is then chosen to leap until the new step does not cross any membranes.

However, random walkers are effectively repulsed away from the membrane since the step encountering the membrane is canceled by the algorithm. This repulsion effect close to the membrane in a thickness of the step size $\delta s$ introduces the bias of diffusivity transverse and parallel to the membrane, as discussed below.

\subsection{Bias in the pore size estimation caused by equal-step-length random leap}
\label{sec:app-erl-radial}
\begin{figure*}[bt!]
\centering
	\includegraphics[width=0.8\textwidth]{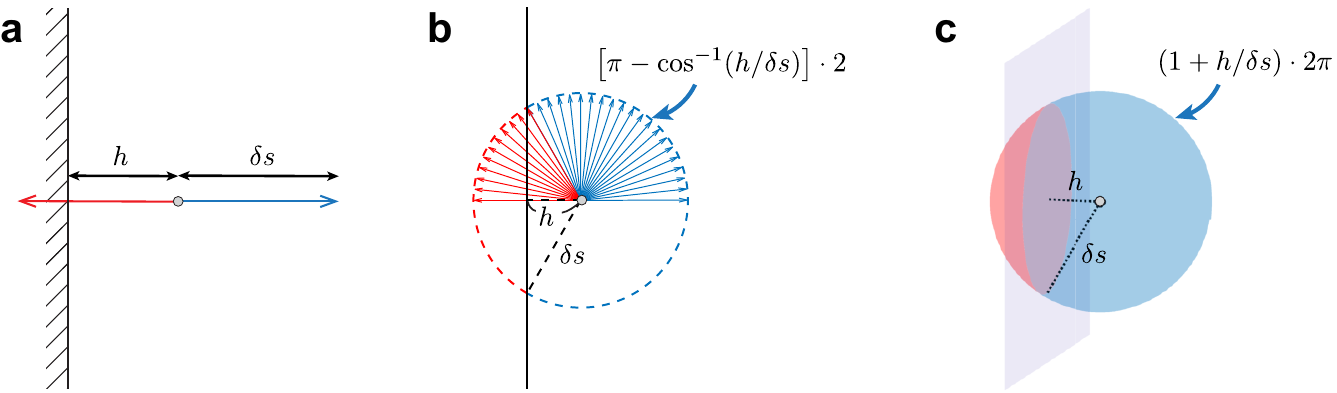}
	\caption{\textbf{The solid angle of the hopping orientation, along which particles do not encounter the membrane (blue).} The step size $\delta s$ and the distance $h$ to the membrane ($h\leq\delta s$) are shown in {\bf a} 1{\it d}, {\bf b} 2{\it d}, and {\bf c} 3{\it d}. The figure is adapted from \citep{fieremans2018cookbook} with the permission from Elsevier.
	}
	\label{fig:solid-angle}
\end{figure*}

\begin{figure}[bt!]
\centering
	\includegraphics[width=0.45\textwidth]{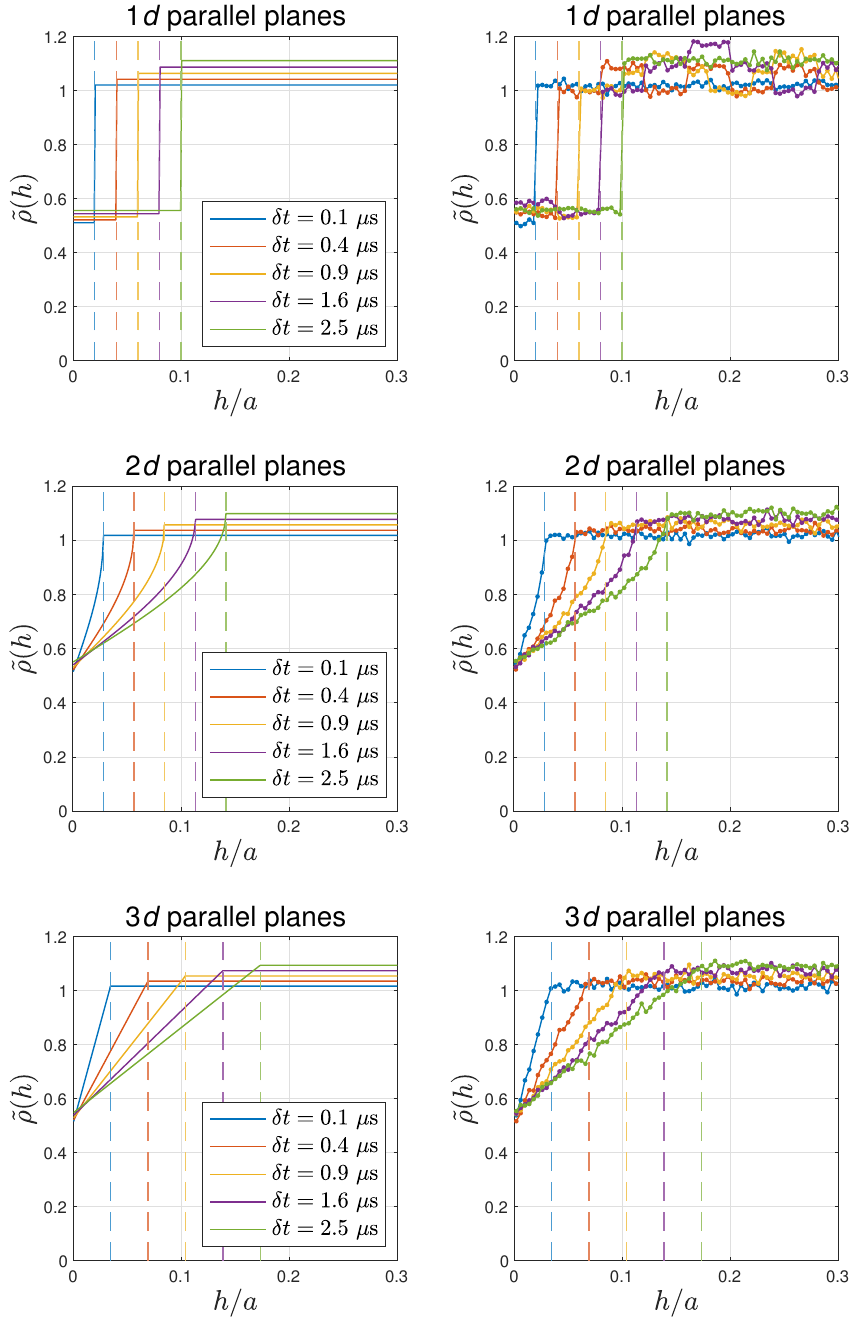}
	\caption{\textbf{Inhomogeneous particle density close to the membrane due to equal-step-length random leap.}
	For elastic collisions, random walkers can not only leap away from the membrane but also leap toward and be reflected by the membrane. However, for the equal-step-length random leap, only the steps jumping away from the boundary are allowed; this repulsive effect results in lower particle density close to the membrane in a thickness of the step size $\delta s$ (indicated by vertical lines). The theoretical prediction of this effect (left column) is consistent with the simulation result (right column) between two parallel planes in a distance $a$ in 1{\it d}, 2{\it d} and 3{\it d}.}
	\label{fig:erl-bias-density}
\end{figure}

To understand the origin of the bias caused by ERL, we firstly discuss the particle density between two parallel planes, composed of two impermeable membranes with a spacing $a$. Due to the time-reversal symmetry of diffusion, the particle density $\rho(h)$ with a distance $h$ to the membrane is proportional to the fraction $p_\Omega(h)$ of the hopping orientation, along which particles do not encounter the membrane (blue solid angle in \figref{fig:solid-angle}):
\begin{align} \label{eq:solid-angle}
    p_\Omega(h\leq\delta s) &= 
    \begin{cases} 
        \frac{1}{2} & d=1 \,,\\
        1-\frac{1}{\pi}\cos^{-1}\left(\frac{h}{\delta s}\right) & d=2 \,, \\
        \frac{1}{2}\left(1+\frac{h}{\delta s}\right) & d=3 \,,
    \end{cases}\\ \notag
    p_\Omega(h>\delta s) &= 1\,.
\end{align}
The fact that $\rho(h)\propto p_\Omega(h)$ in ERL is demonstrated by simulations of diffusion between two parallel planes in 1{\it d}, 2{\it d}, and 3{\it d} in \figref{fig:erl-bias-density}, where the density $\tilde{\rho}(h)$ is normalized by the number of random walkers in simulations, and the step size $\delta s$ in \eqref{eq:step-size} is tuned by varying the time step $\delta t=0.1-2.5$ $\mu$s with the intrinsic diffusivity $D_0 = 2$ $\mu$m$^2$/ms.

The knowledge of the particle density further enables us to estimate the bias in the pore size estimation due to ERL. Considering the diffusion between two parallel planes at $x=\pm a/2$, the diffusion signal measured by using narrow-pulse monopolar sequence is given by \citep{Callaghan1993book}
\begin{equation} \label{eq:S-qt}
    S(q,t) = \int \rho_0(x) \, {\cal{G}}(x,x';t)\, e^{-iq\cdot(x-x')} \text{d}x\,\text{d}x'\,,
\end{equation}
where $q$ is the diffusion wave vector normal to the plane surface, $\rho_0(x)$ is the particle density at $t=0$, and $\cal G$ is the diffusion propagator, i.e., the probability of a spin at $x$ diffusing to $x'$ during the time $t$. 

In MC simulations, on the one hand, we initialize a homogeneous particle density at $t=0$, i.e., $\rho_0(x)=$ const if $x\in[-a/2, a/2]$. On the other hand, at long times ($t\gg t_D = a^2/D_0$), the random walker loses its memory of the initial position and has an equal probability of being anywhere between two planes \citep{Callaghan1993book}:
\begin{equation} \notag
    {\cal G}(x,x';t\gg t_D) \simeq \rho(x) \propto p_\Omega\left(h(x)\right)\,,
\end{equation}
where $p_\Omega(h)$ is given by \eqref{eq:solid-angle} with the distance $h(x)=a/2-|x|$. Substituting into \eqref{eq:S-qt}, we obtain
\begin{equation} \label{eq:S-qt-inf}
    S(q,t\gg t_D)\propto \rho_0(q) \cdot p_\Omega(q)\,,
\end{equation}
where $\rho_0(q)$ and $p_\Omega(q)$ are the corresponding Fourier transform quantities:
\begin{equation} \label{eq:rho-0-q}
    \rho_0(q) = \text{sinc}\left(\frac{qa}{2}\right)\,,
\end{equation}
and
\begin{equation} \label{eq:p-omega-1d}
    p_\Omega (q) = \frac{1}{2}\left[ \text{sinc}\left(\frac{qa}{2}\right) + \text{sinc}\left(q\cdot\left(\frac{a}{2}-\delta s\right)\right) \right]\,,
\end{equation}
for 1-dimensional parallel planes, and
\begin{multline} \label{eq:p-omega-2d}
    p_\Omega(q) = 
    \frac{1}{2}\left[ \text{sinc}\left(\frac{qa}{2}\right) \cdot (J_0(q\delta s)+1) \right.\\
    \left. - \frac{\cos(qa/2)}{qa/2}\cdot H_0(q\delta s)  \right]\,,
\end{multline}
for 2-dimensional parallel planes with $H_0(\cdot)$ the Struve function and $J_0(\cdot)$ the Bessel function of the first kind, and
\begin{equation} \label{eq:p-omega-3d}
    p_\Omega(q) = \frac{1}{2}\left[ \text{sinc}\left(\frac{qa}{2}\right) +  \text{sinc}\left(\frac{q\delta s}{2}\right)\cdot \frac{\sin\left(q\cdot(a-\delta s)/2\right)}{qa/2}\right]\,,
\end{equation}
for 3-dimensional parallel planes. Substituting Eqs.~(\ref{eq:rho-0-q}), (\ref{eq:p-omega-1d}), (\ref{eq:p-omega-2d}) and (\ref{eq:p-omega-3d}) into \eqref{eq:S-qt-inf} and applying Taylor series for $q\delta s \ll qa\ll 1$, we have
\begin{equation} \notag
    -\ln \left(\frac{S}{S_0}\right)_{t\gg t_D} \simeq 
    \begin{cases} 
        \frac{1}{12}q^2\cdot \left(a^2-a\delta s\right) & d=1\,,\\
        \frac{1}{12}q^2\cdot \left(a^2-\frac{2}{\pi}a\delta s\right) & d=2\,,\\
        \frac{1}{12}q^2\cdot \left(a^2-\frac{1}{2}a\delta s\right) & d=3\,,
    \end{cases}
\end{equation}
where $S_0\equiv S|_{q\to0}$.

\begin{figure}[bt!]
\centering
	\includegraphics[width=0.45\textwidth]{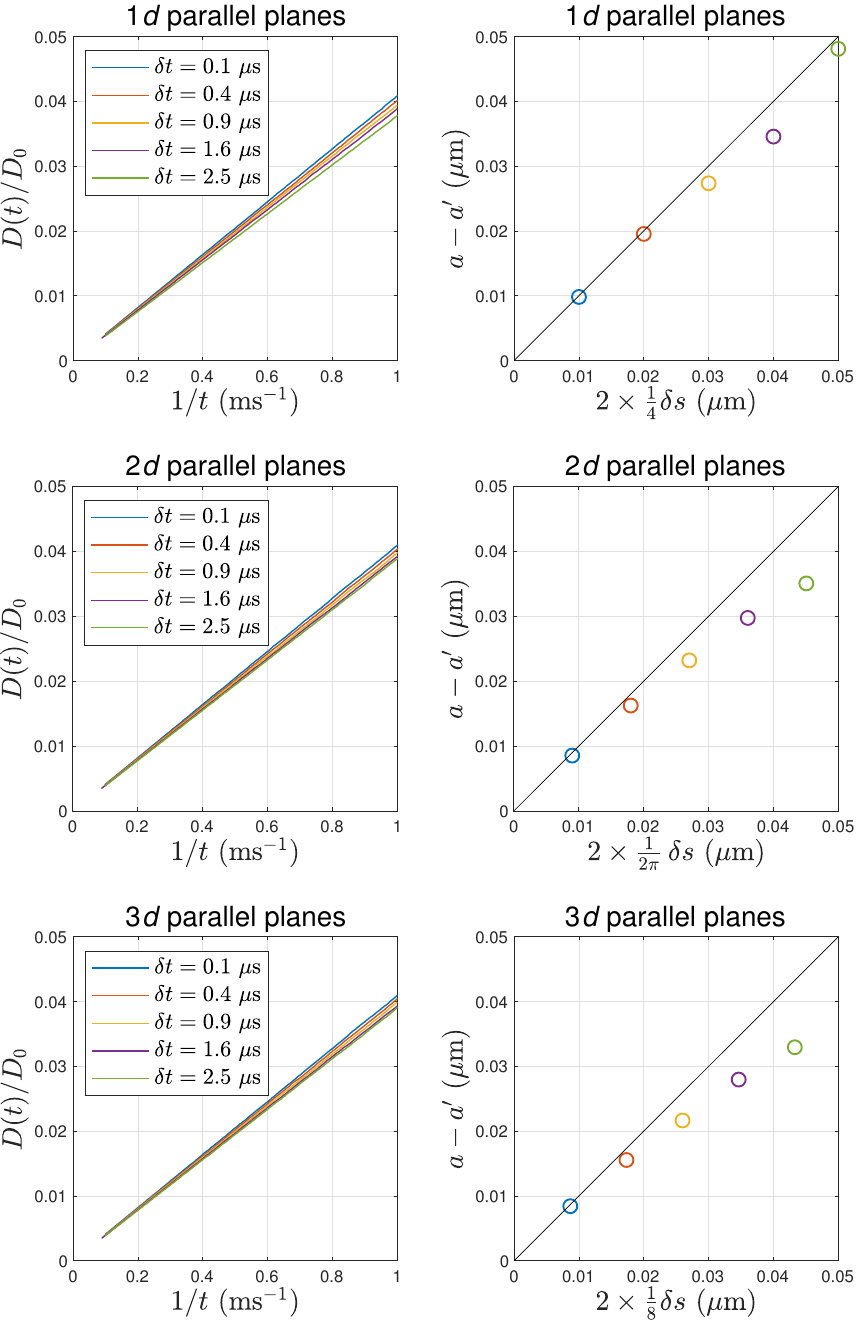}
	\caption{\textbf{The bias of diffusivity transverse to membranes due to equal-step-length random leap.}
	The inhomogeneous particle density due to equal-step-length random leap causes the bias in diffusivity $D(t)$ transverse to membranes and subsequently the pore shape estimation, for example, the distance between two parallel planes. The ground truth of the planes distance is $a=1$ $\mu$m, and its biased estimation based on $D(t)$ is $a'$. This bias $a-a'$ is roughly proportional to the step size $\delta s$.}
	\label{fig:erl-bias-size}
\end{figure}

Using the definition of the apparent diffusivity, $D\equiv-\frac{1}{b}\ln\frac{S}{S_0}$ with $b= q^2t$ in narrow pulse, the diffusivity transverse to parallel planes in simulations of ERL is given by $D(t) = (a')^2/(12 t)$, where
\begin{equation} \label{eq:erl-radial}
    \frac{a-a'}{2}\equiv \delta r \simeq 
    \begin{cases} 
        \frac{1}{4}\delta s & d=1\,,\\
        \frac{1}{2\pi}\delta s & d=2\,,\\
        \frac{1}{8}\delta s & d=3\,.
    \end{cases}
\end{equation}
Comparing with the unbiased solution $D(t)=a^2/(12 t)$ \citep{Callaghan1993book}, the bias of the pore size estimation due to ERL can be considered as an effective pore shrinkage, and the $\delta r$ serves as the shrinkage distance from the membrane. To demonstrate \eqref{eq:erl-radial}, we performed MC simulations of diffusion between two parallel planes in 1{\it d}, 2{\it d}, and 3{\it d} (implemented by using ERL), as shown in \figref{fig:erl-bias-size}, where the step size $\delta s$ is tuned by varying the time step $\delta t$, as in \figref{fig:erl-bias-density}.

\subsection{Bias in the diffusivity parallel to the membrane caused by equal-step-length random leap}
\label{sec:erl-axial}
Considering the diffusion in 3{\it d} parallel to a membrane and its simulation implemented with ERL, the random walker close to the membrane has a higher probability to leap parallel to than perpendicular to the membrane due to the repulsive effect of ERL in \ref{sec:app-erl-radial}. As a result, the diffusion displacement as well as the diffusivity $D_\parallel(t)$ parallel to the membrane are slightly overestimated.

Given that a random walker close to a membrane with a distance $h\leq\delta s$ leaps into a direction of a polar angle $\theta$ and an azimuthal angle $\phi$ (spherical coordinate with zenith direction along $x$-axis), its diffusion displacements parallel to the membrane (along, e.g., $z$-axis) is $\delta z=\delta s\sin\theta\cos\phi$. Averaging over all possible directions allowed by the ERL, the second order cumulant of $\delta z$ is given by
\begin{align*}
\langle \delta z^2\rangle_\Omega &= \int_{\cos^{-1}(\frac{h}{\delta s})}^0 \int_0^{2\pi}(\delta s\sin\theta\cos\phi)^2 \,\frac{\sin\theta \,\text{d}\theta}{1+\frac{h}{\delta s}}\, \frac{\text{d}\phi}{2\pi}\\
&= \tfrac{1}{6}(2\delta s^2+h\delta s-h^2)\,,\quad h\leq \delta s\,.
\end{align*}
Further averaging over the thickness $\delta s$ surrounding the membrane with the consideration of the inhomogeneous density $\rho(h)\propto p_\Omega(h)$ in 3{\it d} in \eqref{eq:solid-angle}, we obtain the second order cumulant of $\delta z$, given by
\begin{equation*}
\langle \delta z^2 \rangle_{h\leq\delta s} = \frac{\int_0^{\delta s}\langle \delta z^2\rangle_\Omega\cdot p_\Omega \,\text{d}h}{\int_0^{\delta s} p_\Omega\, \text{d}h}=\frac{13}{36} \delta s^2\,,
\end{equation*}
leading to a distinct diffusivity parallel to the membrane for walkers close to the membrane:
\begin{equation*}
D_0'=\frac{\langle \delta z^2 \rangle_{h\leq\delta s}}{2\delta t}=\frac{13}{12}D_0\,,
\end{equation*}
where the step size in \eqref{eq:step-size} in 3{\it d} is applied.

Random walkers surrounding the membrane ($h\leq\delta s$) have a diffusivity $D_0'$ parallel to the membrane and a (density-weighted) volume fraction 
\begin{align*}
    f' &= \int_0^{\delta s} \rho(h) \cdot \frac{S}{V}\,\text{d}h \\
    &\simeq \int_0^{\delta s} p_\Omega(h)\cdot \frac{S}{V}\,\text{d}h = \frac{3}{4}\cdot\frac{S}{V}\delta s\,,
\end{align*}
where $\frac{S}{V}$ is the surface-to-volume ratio. Similarly, random walkers away from the membrane have a diffusivity $D_0$ parallel to the membrane and a volume fraction $f=1-f'$. Their volume-weighted sum yields the overall diffusivity parallel to the membrane, given by
\begin{align} \notag
D_\parallel&=f\cdot D_0 + f'\cdot D_0'\\ \label{eq:erl-axial}
&\simeq D_0\cdot\left(1+\frac{1}{16}\frac{S}{V}\delta s\right) > D_0\,,
\end{align}
where the correction term is negligible when a small step size is applied. 

\begin{figure}[t]
\centering
	\includegraphics[width=0.45\textwidth]{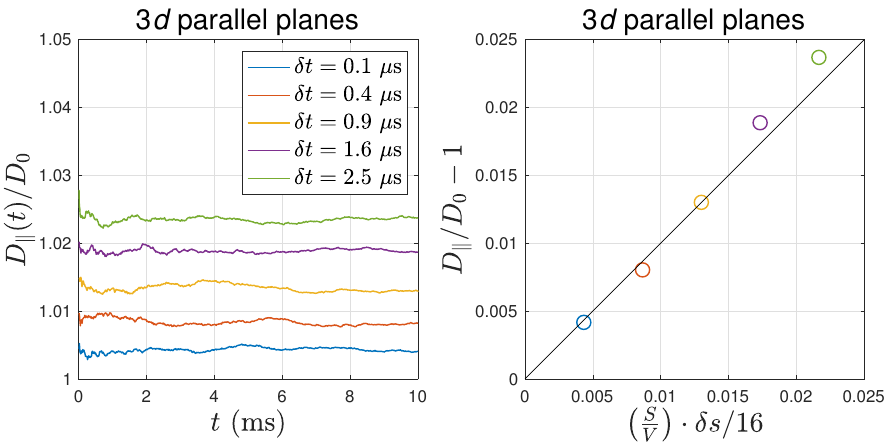}
	\caption{\textbf{The bias of diffusivity parallel to membranes due to equal-step-length random leap.}
	The inhomogeneous particle density due to equal-step-length random leap causes the bias in diffusivity $D_\parallel$ parallel to the membrane (e.g., two parallel planes in 3{\it d}). This bias is roughly proportional to the step size $\delta s$ and closely related to the surface-to-volume ratio $S/V$ of the geometry.}
	\label{fig:erl-bias-axial}
\end{figure}

The above \eqref{eq:erl-axial} was demonstrated by performing MC simulations of diffusion between two parallel planes in 3{\it d} (implemented by using ERL), as shown in \figref{fig:erl-bias-axial}, where the step size $\delta s$ is tuned by varying the time step $\delta t$, as in \figref{fig:erl-bias-density}.

\section{Unbiased simulation of particle-membrane interaction --- rejection sampling}
\label{sec:app-rs}
\renewcommand{\theequation}{B.\arabic{equation}}
\setcounter{equation}{0}
\renewcommand{\thefigure}{B.\arabic{figure}}
\setcounter{figure}{0}

Another alternative of elastic collision is rejection sampling \citep{ford1997simulation,waudby2011gpu,nguyen2018gpu,palombo2018leaflet}: a step crossing a membrane is canceled, and the random walker does not move for the step. This simple approach properly maintains a homogeneous particle density in simulations of diffusion in the substrate composed of impermeable membranes \citep{szymczak2003rsampling} and can be generalized to the case of permeable membranes as follows.

\subsection{Generalization of rejection sampling to the case of permeable membranes}
\label{sec:app-rs-perm}
Here, we use the theoretical framework similar to that in Appendix 1 and Figure 7 of \citep{fieremans2010karger}: Given that, in 1{\it d}, a permeable membrane is positioned at $x_M=0$, and the intrinsic diffusivity at both sides is $D_0$, the probability of a random walker showing at position $x_P$ on the right side of the membrane at time $t+\delta t$, such that $x_P-x_M<\delta s$, is related to the probabilities of previous steps at time $t$, i.e.,
\begin{equation} \label{eq:rs-pdf}
    {\cal P}(x_P,t+\delta t) = \frac{1}{2}{\cal P}(x_P+\delta s,t) + \frac{P_T}{2}{\cal P}(x_P-\delta s,t) + \frac{1-P_T}{2}{\cal P}(x_P,t)\,,
\end{equation}
where $\delta t$ is the time step, $\delta s$ is the step size in \eqref{eq:step-size} in 1{\it d}, $P_T$ is the permeation probability, and ${\cal P}(\cdot)$ is the probability density function (PDF) of the particle population at a given position and time. Notably, the third right-hand-side term in \eqref{eq:rs-pdf} corresponds to the canceled step without update due to the rejection sampling. This minor difference from the Eq. [40] in \citep{fieremans2010karger} leads to slightly different result in the permeation probability.


The PDF obeys the diffusion equation,
\begin{equation*}
    D_0\partial_x^2{\cal P} = \partial_t{\cal P}\,,
\end{equation*}
and the boundary condition at the permeable membrane of permeability $\kappa$ is given by
\begin{equation*}
    \kappa\left({\cal P}^- - {\cal P}^+\right) = -D_0 \partial_x {\cal P}^-\,, \quad D_0 \partial_x {\cal P}^- = D_0 \partial_x{\cal P}^+\,.
\end{equation*}
Substituting into the Taylor expansion of \eqref{eq:rs-pdf} at $x_M$ and $t$ and ignoring higher order terms, we obtain the permeation probability specifically for the implementation of rejection sampling:
\begin{equation*}
    \frac{P_T}{1-P_T} = \frac{\kappa\delta s}{D_0}\,,
\end{equation*}
similar to the functional form in \eqref{P-Els}, except that $P_T$ is related to the step size, rather than the particle's distance to the membrane. Likewise, the above expression of $P_T$ can be generalized in 2{\it d} and 3{\it d} by using Eqs.~(\ref{eq:perm-prob}) and (\ref{eq:kappa-correction}) in \ref{sec:app-permeability}.
\rem{I thought in 2010 Appendix we derived the formula with the distance to membrane, not with $\delta s$. I think you should first give that formula here (basically Eq C.1), and then provide the rationale why you want to approximate it by its average, adn then say where it is derived. 1. For elastic collision, that is distance to the membrane. For rejection sampling, it is the step size related to the probability. The Eq. B.1 is different from that in Fieremans et al. 2010. The difference is emphasized now.}

\begin{figure}[ht!]
\centering
	\includegraphics[width=0.45\textwidth]{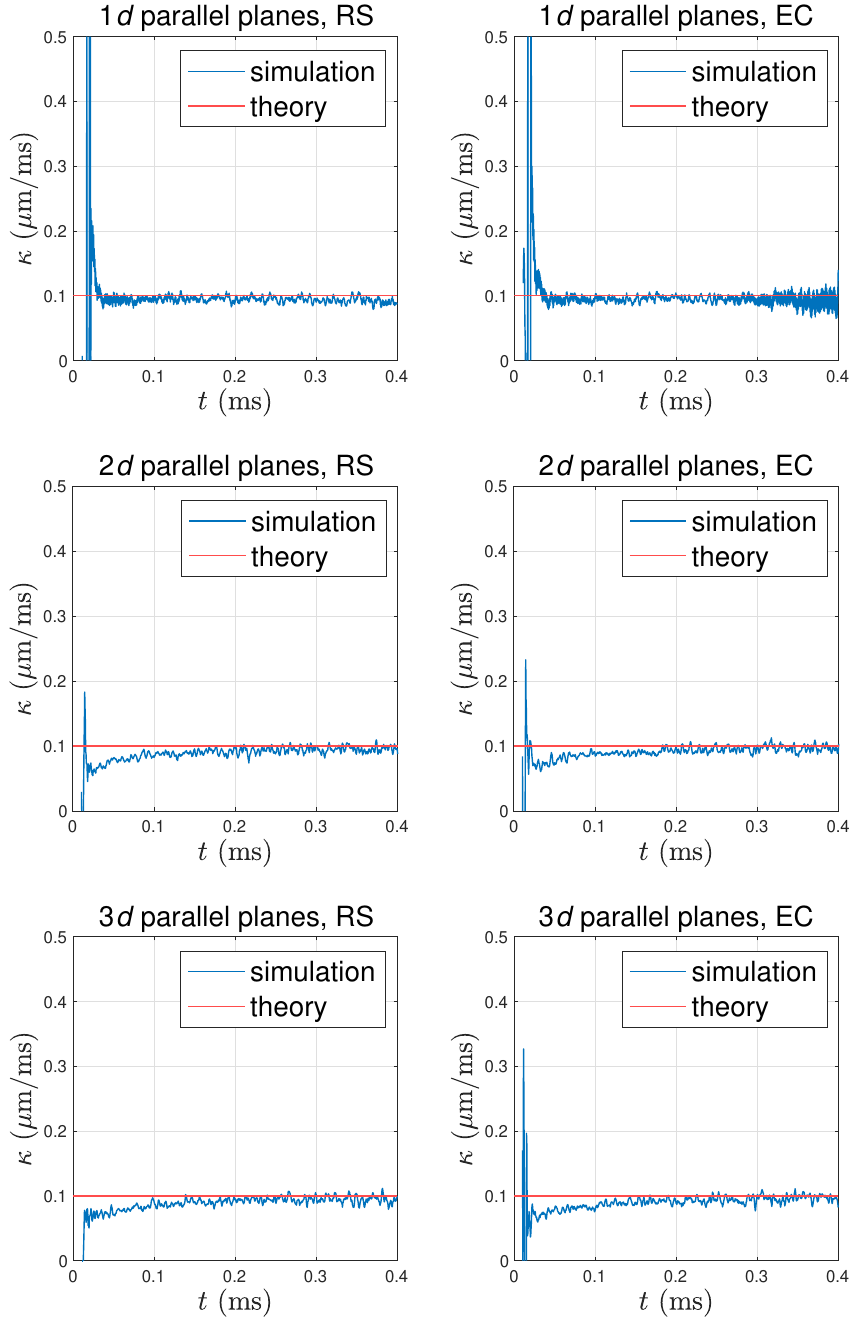}
	\mnote{R3.4\\R3.6}
	\caption{\textbf{Permeability calculated based on the particle density around the membrane, which interacts with random walkers through rejection sampling (RS) or elastic collision (EC).} 
	When a random walker encounters a membrane of permeability $\kappa$, the permeation probability is determined by \eqref{eq:perm-prob-general}, whose applicability for RS and EC is demonstrated by permeability calculated based on the simulated particle density around two parallel planes (membranes) separated by a distance $a=1$ $\mu$m in 1$d$, 2$d$ and 3$d$. To create a density impulse at time $t=0$, random walkers are initialized at $x=0$ with membranes at $x_M=\pm a/2$. 
	The calculated permeability is consistent with the theoretical value input in \eqref{eq:perm-prob-general}. To show the trend of data without distorting the tendency, the curve of $\kappa$ over $t$ is smoothed by applying a Savitzky-Golay filter of 11-point cubic polynomial.
	}
	\label{fig:rs-ec-permeability}
\end{figure}

\mpar{R3.4\\R3.6}\new{
It has been shown that both rejection sampling and elastic collision maintain homogeneous particle density around membranes \citep{szymczak2003rsampling}. Furthermore, to demonstrate the permeation probability in \eqref{eq:perm-prob-general} for both interactions, we performed MC simulations of diffusion between two parallel planes separated by a distance $a=1$ $\mu$m in 1$d$, 2$d$, and 3$d$ (implemented by using rejection sampling or elastic collision). We created a localized density source of Dirac delta function at time $t=0$, i.e., $\rho_0(x) = \delta_D(x)$, half-way between two permeable membranes of permeability $=0.1$ $\mu$m/ms at $x_M=\pm a/2$. Then we calculated the permeability $\kappa$ based on the particle density $\rho(x)$ around membranes \citep{powles1992simulation,fieremans2008simulation}:
\begin{equation} \notag
    -D_0^+\cdot\partial_x\, \rho(x_M^+) 
    = -D_0^-\cdot\partial_x\, \rho(x_M^-) 
    = \kappa\cdot\left[\rho(x_M^-) - \rho(x_M^+)\right]\,,
\end{equation}
where $D_0^+$ and $D_0^-$ were diffusivities on the right and left side of the membrane, and $x_M^+$ and $x_M^-$ were positions at the right and left side of the membrane. Here we applied $2\times10^7$ random walkers diffusing over 1000 steps with a step duration $\delta t=4\times10^{-4}$ ms and a step size given by \eqref{eq:step-size}, where $D_0 = D_0^+ = D_0^- = 0.5$ $\mu$m$^2$/ms. 

For both rejection sampling and elastic collision, the permeability calculated based on the particle density is consistent with the theoretical value input in \eqref{eq:perm-prob-general} (\figref{fig:rs-ec-permeability}). Ideally, the calculated permeability is independent of diffusion time, and the spurious permeability time-dependence at short time is, to the best of our understanding, caused by the discretization error due to the small number of steps, when too few particles reach the membrane and contribute to the flux.
}

\begin{figure}[t]
\centering
	\includegraphics[width=0.45\textwidth]{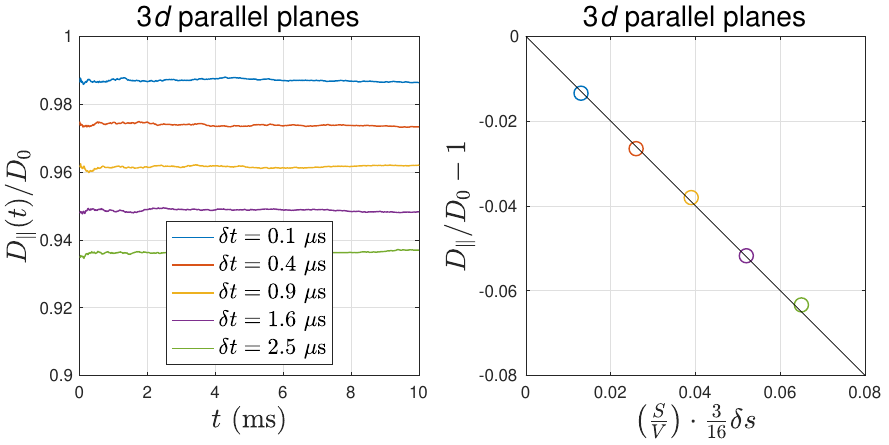}
	\caption{\textbf{The bias of diffusivity parallel to membranes due to rejection sampling.}
	The algorithm of rejecting steps toward a membrane due to rejection sampling causes the bias in diffusivity $D_\parallel$ parallel to the membrane (e.g., two parallel planes in 3{\it d}). This bias is roughly proportional to the step size $\delta s$ and closely related to the surface-to-volume ratio $S/V$ of the geometry.}
	\label{fig:rs-bias-axial}
\end{figure}

\subsection{Bias in the diffusivity parallel to the membrane caused by rejection sampling}
In this section, we will follow the framework in \ref{sec:erl-axial} to evaluate the bias in diffusivity parallel to the membrane due to rejection sampling. Considering that the diffusion around a membrane in 3{\it d} obeys the rejection sampling, the random walker cancels some steps (without updates) toward but not completely transverse to the membrane, leading to smaller diffusion displacement cumulant as well as diffusivity $D_\parallel(t)$ parallel to the membrane.

Given that a random walker close to a membrane with a distance $h\leq\delta s$ leaps into a direction of a polar angle $\theta$ and an azimuthal angle $\phi$ (spherical coordinate with zenith direction along $x$-axis), its diffusion displacements parallel to the membrane (along, e.g., $z$-axis) is $\delta z=\delta s\sin\theta\cos\phi$. Averaging over all possible directions allowed by the rejection sampling, the second order cumulant of $\delta z$ is given by
\begin{align*}
\langle \delta z^2\rangle_\Omega &= \int_{\cos^{-1}(\frac{h}{\delta s})}^0 \int_0^{2\pi}(\delta s\sin\theta\cos\phi)^2 \,\frac{\sin\theta \,\text{d}\theta\,\text{d}\phi}{4\pi}\,\\
&= \frac{\delta s^2}{12}\left[2+3\frac{h}{\delta s}-\left(\frac{h}{\delta s}\right)^3\right]\,,\quad h\leq \delta s\,.
\end{align*}
Further averaging over the thickness $\delta s$ surrounding the membrane with a homogeneous particle density maintained by rejection sampling, we obtain the second order cumulant of $\delta z$, given by
\begin{equation*}
\langle \delta z^2 \rangle_{h\leq\delta s} = \int_0^{\delta s}\langle \delta z^2\rangle_\Omega\cdot p_\Omega \,\frac{\text{d}h}{\delta s}=\frac{13}{48} \delta s^2\,,
\end{equation*}
leading to a distinct diffusivity parallel to the membrane for walkers close to the membrane:
\begin{equation*}
D_0'=\frac{\langle \delta z^2 \rangle_{h\leq\delta s}}{2\delta t}=\frac{13}{16}D_0\,,
\end{equation*}
where the step size in \eqref{eq:step-size} in 3{\it d} is applied.

Random walkers surrounding the membrane ($h\leq\delta s$) have a diffusivity $D_0'$ parallel to the membrane and a volume fraction $f'=\frac{S}{V}\delta s$. Similarly, random walkers away from the membrane have a diffusivity $D_0$ parallel to the membrane and a volume fraction $f=1-f'$. Their volume-weighted sum yields the overall diffusivity parallel to the membrane, given by
\begin{align} \notag
D_\parallel&=f\cdot D_0 + f'\cdot D_0'\\ \label{eq:rs-axial}
&\simeq D_0\cdot\left(1-\frac{3}{16}\frac{S}{V}\delta s\right) < D_0\,,
\end{align}
where the correction term is negligible when a small step size is applied. 

The above \eqref{eq:rs-axial} was demonstrated by performing MC simulations of diffusion between two parallel planes in 3{\it d} (implemented by using rejection sampling), as shown in \figref{fig:rs-bias-axial}, where the step size $\delta s$ is tuned by varying the time step $\delta t$.

\section{Unbiased simulation of membrane permeability}
\label{sec:app-permeability}
\renewcommand{\theequation}{C.\arabic{equation}}
\setcounter{equation}{0}
\renewcommand{\thefigure}{C.\arabic{figure}}
\setcounter{figure}{0}

Here we introduce the theoretical background of unbiased simulations of permeable membranes and provide a first order correction of permeation probability by considering the particle density flux around membranes. The theoretical results extend the applicability of related simulation models and offer a guide to choose simulation parameters. In this section, the particle density at position $x$ is denoted by $\rho(x)$.

\subsection{The physics of the permeability correction: Equal molecular concentration}
The discussion in this section follows the logic of the Appendix B in our previous work \citep{lee2020gray}.
Given that a random walker encounters a membrane of permeability $\kappa$, the  permeation probability $P$ through the membrane is related with the distance $h$ between the random walker and the encountered membrane when $h\leq\delta s$, with $\delta s$ the step size in \eqref{eq:step-size}, $D_0$ the intrinsic diffusivity and $\delta t$ the time step in $d$ dimensional space, as derived in Appendix A of \citep{fieremans2010karger}, Eq.~(43):
\begin{equation} \label{P-Els}
    \frac{P}{1-P}=\frac{2\kappa h}{D_0}\,,
\end{equation}
which is a well-regularized functional form of $P$ even for a highly permeable membrane: As expected, the limit $\kappa\to\infty$ corresponds to the probability $P\to1$.

In actual implementations, to reduce the computational load due to the calculation of the distance $h$ from a random walker to the encountered membrane, the distance $h$ could be approximated by the step size $\delta s$, 
by {\it averaging} the particle density flux $j(h)$ around the membrane over $h\leq \delta s$. To do so, a low probability is assumed ($P\ll1$), such that the denominator in the left-hand-side term in \eqref{P-Els} is about $1$, i.e., $P(h)\simeq 2\kappa_0 h/D_0$, where the permeability is redefined ($\kappa \to \kappa_0$) since the approximate relation is not exact.
Averaging the particle density flux $j(h)$ is equivalent to averaging the permeation probability $P(h)$ because $j\propto P$, leading to
\begin{equation} \notag
    P(h) \to P(\delta s) \simeq \frac{\int_0^{\delta s} P(h) \, \tilde{p}_\Omega(h\leq\delta s)\,\text{d}h }{\int_0^{\delta s} \tilde{p}_\Omega(h\leq\delta s)\,\text{d}h}=
    \frac{\kappa_0 \delta s}{D_0}\cdot C_d \ll 1 \,,
\end{equation}
where $\tilde{p}_\Omega = 1-p_\Omega$ is the fraction of directions encountering the membrane, corresponding to the red solid angle in \figref{fig:solid-angle}, and $p_\Omega$ is defined in \eqref{eq:solid-angle}.

In this case, the permeation probability is given by \eqref{eq:perm-prob},
where its presumption, i.e., $P\ll1$, yields a condition to be satisfied:
\begin{equation*}
    \kappa_0\ll\sqrt{\frac{D_0}{2d\delta t}}\cdot\frac{1}{C_d}\,,
\end{equation*}
indicating that using a short time step $\delta t$ is required in simulations of exchange through a highly permeable membrane (large $\kappa_0$); in this case, the genuine permeability $\kappa$ is close to the input value $\kappa_0$.

For larger $\kappa$, the input $\kappa_0$ would be significantly different from the genuine value $\kappa$ in simulations while extending the approximation of $h$ by $\delta s$. We will show that averaging over $h\leq\delta s$ simply renormalizes the input $\kappa_0$ entering Eq.~(\ref{eq:perm-prob}), leading to a genuine $\kappa$ in \eqref{eq:kappa-correction} given below. 
The idea behind is that averaging over $\delta s$ affects not just the permeation probability but also the particle flux density. We here demand the Fick's first law satisfying the permeation probability in \eqref{eq:perm-prob} and derive a correction factor renormalizing the permeability $\kappa_0 \to \kappa$.


The particle density flux from left to right  ($1\to2$) is  given by
\begin{equation} \label{eq:flux1-2}
    j_{1\to2}\simeq \frac{ (2dC_d)^{-1} \cdot \langle \rho_1\rangle \cdot(S \delta s_1) \cdot P_{1\to2}}{S\cdot \delta t}\,,
\end{equation}
where $S$ is the surface area, $P_{1\to2}$ is the permeation probability from left to right given by \eqref{eq:perm-prob}, $\langle \rho_1\rangle$ is the particle density averaged over the layer (of thickness $\delta s_1$) on the left side of the membrane, and $(2dC_d)^{-1}$ is the mean fraction of the hopping orientation (averaged over $\delta s_1$ as well) along which particles encounter the membrane \citep{fieremans2018cookbook}.

Substituting \eqref{eq:perm-prob} into \eqref{eq:flux1-2} and using $(\delta s_1)^2/\delta t = 2d D_1$, we obtain $j_{1\to2}\simeq \kappa_0\cdot\langle\rho_1\rangle$. Similarly, the particle density flux from right to left side is $j_{2\to1}\simeq -\kappa_0 \cdot \langle\rho_2\rangle$. Then the net particle flux density is given by
\begin{equation} \label{eq:flux-avg-density}
    j = j_{1\to2} + j_{2\to1} \simeq \kappa_0\cdot\left(\langle\rho_1\rangle - \langle\rho_2\rangle \right)\,.
\end{equation}

Given that the particle density {\it right} at left and right surface of the membrane is $\rho_{0,1}$ and $\rho_{0,2}$ (without spatial averaging), the net particle density flux is (by definition of the genuine $\kappa$)
\begin{align} \label{eq:flux-density}
    j &= \kappa \cdot (\rho_{0,1} - \rho_{0,2}) \\ \label{eq:flux-grad}
      &=-D_1\cdot \partial_x \, \rho_{0,1} = -D_2 \cdot \partial_x \, \rho_{0,2}\,,
\end{align}
where $\kappa$ is the genuine permeability that we would like to achieve with simulations, different from the input value $\kappa_0$, and $\partial_x \, \rho_{0,1}$ and $\partial_x \, \rho_{0,2}$ are density gradients {\it right} at left and right surface of the membrane.

Here we average the density $\langle\rho_1\rangle$ (and $\langle\rho_2\rangle$) over the layer on the left (and right) side of the membrane, of thickness $\delta s_1$ (and $\delta s_2$), and equate the flux density in \eqref{eq:flux-avg-density} to that in \eqref{eq:flux-density} to obtain the genuine permeability $\kappa$. 

Approximating the particle density ($\rho_1$, $\rho_2$) variation close to the membrane with a linear function of the distance from the membrane, we have
\begin{equation*} 
    \rho_1(x)\simeq \rho_{0,1}+\partial_x\,\rho_{0,1}\cdot x\,,\quad
    \rho_2(x)\simeq \rho_{0,2}+\partial_x\,\rho_{0,2}\cdot x\,.
\end{equation*}
Considering the fraction $\tilde{p}_\Omega(x)$ of the hopping orientation along which particles encounter the membrane, as shown in Fig. A1 in \citep{fieremans2018cookbook}, the particle density averaged within the thickness of step size is given by
\begin{subequations}  \label{eq:density-taylor} \begin{align} 
    \langle \rho_1\rangle &= \frac{\int_{-\delta s_1}^0 \rho_1(x) \tilde{p}_\Omega(x)\, \text{d} x}{\int_{-\delta s_1}^0 \tilde{p}_\Omega(x) \,\text{d} x} = \rho_{0,1} - \partial_x\,\rho_{0,1} \cdot \delta s_1 \cdot \frac{C_d}{2}\,,\\
    \langle \rho_2\rangle &= \frac{\int_0^{\delta s_2} \rho_2(x) \tilde{p}_\Omega(x) \,\text{d} x}{\int_0^{\delta s_2} \tilde{p}_\Omega(x) \,\text{d} x} = \rho_{0,2} + \partial_x\,\rho_{0,2} \cdot \delta s_2 \cdot \frac{C_d}{2}\,,
\end{align} \end{subequations}
where 
\begin{equation} \notag
    \tilde{p}_\Omega(x) = 
    \begin{cases}
        1-p_\Omega(|x|\leq\delta s_1) & x<0\,,\\
        1-p_\Omega(|x|\leq\delta s_2) & x\geq0\,,
    \end{cases}
\end{equation}
and $p_\Omega$ is defined in \eqref{eq:solid-angle}.

Substituting Eqs.~(\ref{eq:flux-avg-density}) and (\ref{eq:flux-grad})--(\ref{eq:density-taylor}) into \eqref{eq:flux-density} yields
\begin{equation} \label{eq:kappa-correction}
    \kappa = \frac{\kappa_0}{1-\alpha} > \kappa_0\,,
\end{equation}
where
\begin{align} \notag 
    \alpha &= \frac{1}{2}\kappa_0\left( \frac{\delta s_1}{D_1} + \frac{\delta s_2}{D_2}\right)\cdot C_d\\ \label{eq:kappa-factor-prob}
    &=\frac{P_{1\to2} + P_{2\to1}}{2}\,.
\end{align}

Interestingly, the correction factor $\alpha$ is the permeation probability averaged for both directions, i.e., $\alpha \in [0,1]$. Therefore, the genuine permeability $\kappa$ in Monte Carlo simulations of any dimension $d$ is {\it always larger} than the input value $\kappa_0$, as in \eqref{eq:kappa-correction}, where the correction factor $\alpha$ is essential especially when simulating the diffusion across a highly permeable membrane. To minimize $\alpha$ and reduce the bias, a smaller time-step and larger intrinsic diffusivity should be used.

Practically, to simulate a membrane of permeability $\kappa$, we have to tune the input permeability $\kappa_0$ for the permeation permeability in \eqref{eq:perm-prob} based on
\begin{equation*}
    \kappa_0 = \frac{\kappa}{1+\kappa\cdot(\alpha/\kappa_0)}\,,
\end{equation*}
where the right-hand side is independent of $\kappa_0$ due to \eqref{eq:kappa-factor-prob}.
Substituting the above relation into \eqref{eq:perm-prob} yields the corrected permeation probability in \eqref{eq:perm-prob-general}.
The above correction ensures the genuine permeability $\kappa$ in simulations.

Furthermore, the corrected permeation probability in \eqref{eq:perm-prob-general} should still be $\ll1$, leading to the following constraint, as a guidance of choosing simulation parameters:
\begin{equation*}
    \kappa\ll\sqrt{\frac{2}{d\delta t}}\cdot\frac{1}{C_d}\cdot\frac{\sqrt{D_1D_2}}{\left|\sqrt{D_1}-\sqrt{D_2}\right|}\,.
\end{equation*}
In other words, \eqref{eq:perm-prob-general} works particularly well for a small time-step $\delta t$, large intrinsic diffusivities $\sqrt{D_1 D_2}$, and similar intrinsic diffusivities between compartments ($D_1\simeq D_2$).

\begin{figure}[bt!]
\centering
	\includegraphics[width=0.45\textwidth]{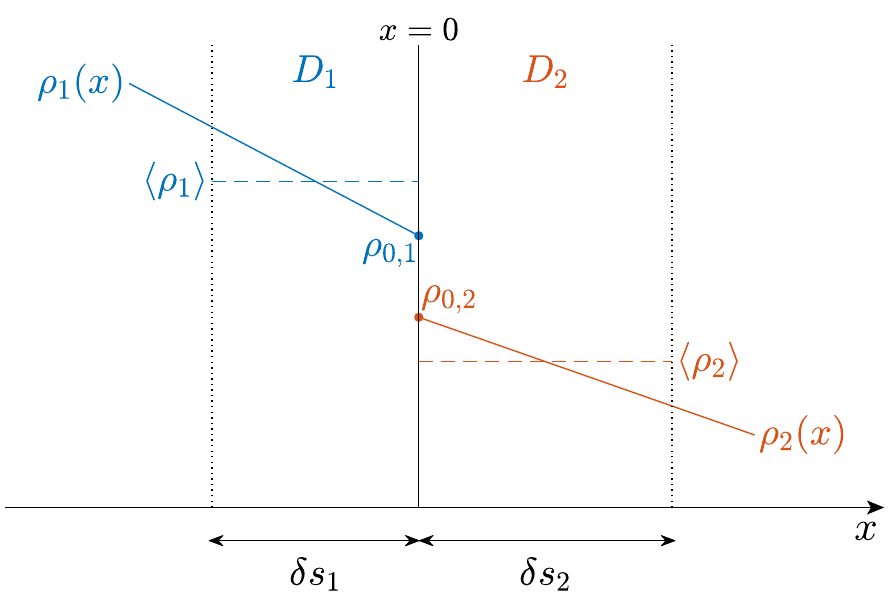}
	\caption{\textbf{Non-zero density flux close to the membrane leads to the bias in membrane permeability.}
	The particle density across a permeable membrane has an offset due to the finite permeability. The input permeability $\kappa_0$ is related to the difference of particle density ($\langle \rho_1\rangle$, $\langle \rho_2\rangle$) averaged within the thickness of the hopping step size ($\delta s_1$, $\delta s_2$); however, the genuine permeability $\kappa$, to be achieved in simulations, is related to the difference of the particle density ($\rho_{0,1}$, $\rho_{0,2}$) at the membrane. As a result, the genuine permeability $\kappa$ is always larger than the input value $\kappa_0$. To clarify, only density flux to the right is plotted in this figure. In actual simulations, the balance of the density flux in both directions yields a homogeneous particle density in each compartment.}
	\label{fig:bias-density}
\end{figure}

\subsection{General case: Different spin concentration at both sides of the membrane}
In the previous section, the medium is assumed to have the same spin concentration in all compartments. However, a lower spin concentration is expected for some tissue microstructure, such as myelin water. To generalize for different spin concentrations in each compartment, the permeation probability in \eqref{eq:perm-prob} is re-written as
\begin{subequations} \label{eq:gen-perm-prob} \begin{align}
    P_{1\to2}&\simeq \frac{\kappa_0\delta s_1}{D_1}\cdot C_d\cdot\left(\frac{c_2}{c_1}\right)^\lambda\,,\\
    P_{2\to1}&\simeq \frac{\kappa_0\delta s_2}{D_2}\cdot C_d\cdot\left(\frac{c_2}{c_1}\right)^{\lambda-1}\,,
\end{align} \end{subequations}
where $c_1$ and $c_2$ are spin concentrations over the left and right sides of the membrane, and $\lambda\in[0,1]$ is an exponent determined later. It is worthwhile to notice that the probability ratio $P_{1\to2}/P_{2\to1}=c_2\sqrt{D_2}/c_1\sqrt{D_1}$ is maintained to ensure the particle density equilibrium for all diffusion times.

Similar to the derivation in previous section, substituting \eqref{eq:gen-perm-prob} into \eqref{eq:flux1-2} and calculating $j_{1\to2}$ and $j_{2\to1}$, we obtain
\begin{align} \notag
    j &= j_{1\to2}+j_{2\to1}\\ \label{eq:gen-flux-avg-density}
      &\simeq \kappa_0\cdot\left[\langle\rho_1\rangle\cdot\left(\frac{c_2}{c_1}\right)^\lambda - \langle\rho_2\rangle\cdot\left(\frac{c_2}{c_1}\right)^{\lambda-1}\right]\,.
\end{align}
Considering the ratio $c_2/c_1$ of spin concentrations, the net particle flux density is given by
\begin{align} \label{eq:gen-flux-density}
    j &= \kappa\cdot\left[\rho_{0,1}\cdot\left(\frac{c_2}{c_1}\right)^\lambda - \rho_{0,2}\cdot\left(\frac{c_2}{c_1}\right)^{\lambda-1}\right]\\ \label{eq:gen-flux-grad}
      &= -D_1\partial_x\,\rho_{0,1} = -D_2\partial_x\,\rho_{0,2}\,,
\end{align}
where the unbiased permeability $\kappa$ is re-defined accordingly.

Substituting Eqs.~(\ref{eq:density-taylor}), (\ref{eq:gen-flux-avg-density}) and (\ref{eq:gen-flux-grad}) into \eqref{eq:gen-flux-density} yields
\begin{equation} \label{eq:gen-kappa-correction}
    \kappa = \frac{\kappa_0}{1-\alpha_\lambda}\,,
\end{equation}
where
\begin{align} \notag
    \alpha_\lambda &= \frac{1}{2}\kappa_0\left[\frac{\delta s_1}{D_1}\cdot\left(\frac{c_2}{c_1}\right)^\lambda + \frac{\delta s_2}{D_2}\cdot\left(\frac{c_2}{c_1}\right)^{\lambda-1}\right]\cdot C_d\\ \label{eq:gen-kappa-factor-prob}
    &=\frac{P_{1\to2}+P_{2\to1}}{2}\,.
\end{align}
The choice of $\lambda$ is essential not only for the generalization of permeability definition, as in Eqs.~(\ref{eq:gen-perm-prob})--(\ref{eq:gen-flux-density}), but also for the permeability bias in \eqref{eq:gen-kappa-correction}. On the one hand, to maintain the same permeability definition for all membranes in the medium, we can fix $\lambda$ as a constant, e.g., 0, 1/2 or 1. On the other hand, to minimize the correction factor $\alpha_\lambda$, we can chose the $\lambda$ based on
\begin{align*}
    \lambda &= \underset{\lambda\in[0,1]}{\text{arg min}}(\alpha_\lambda)\\
            &= \left(1+\frac{c_2\sqrt{D_2}}{c_1\sqrt{D_1}}\right)^{-1}\,,
\end{align*}
which is well-regularized for even extreme cases, e.g., $c_1=0$ or $c_2=0$.

\subsection{Alternative approach for simulations of permeable membrane}
Instead of assigning a nominal permeability $\kappa$ for a permeable membrane, \citet{baxter2013simulation} defined the permeation probability based on the spin concentration ($c_1$, $c_2$) and intrinsic diffusivity ($D_1$, $D_2$) over the left and right side of the membrane:
\begin{subequations} \label{eq:alt-perm-prob1-2} \begin{align}
    P_{1\to2} &= \frac{c_2\sqrt{D_2}}{c_1\sqrt{D_1}}\,,\\
    P_{2\to1} &= 1\,,
\end{align} \end{subequations}
where the left side compartment 1 is a ``high-flux medium'', compared with the right side compartment 2, i.e., $c_1\sqrt{D_1}\geq c_2\sqrt{D_2}$. 
This approach has been applied in simulations of, e.g., the exchange between intra-/extra-axonal water an myelin water \citep{harkins2016myelin} and the water exchange between intra-axonal cytoplasm and mitochondria \citep{lee2020axial}.
It seems that this method introduces neither adjustable parameters for membrane permeability nor particle density transition over the membrane; however, this is true only for infinitely small time-step $\delta t$. For finite $\delta t$, a $\delta t$-dependent permeability may emerge in simulations.

The derivation of this extra permeability is similar to those in previous sections. Substituting Eqs.~(\ref{eq:density-taylor}) and (\ref{eq:alt-perm-prob1-2}) into \eqref{eq:flux1-2}, the particle density flux for both directions is given by
\begin{align*}
    j_{1\to2} &\simeq \kappa_0' \cdot \left(\rho_{0,1}-\partial_x\,\rho_{0,1}\cdot \delta s_1\cdot \frac{C_d}{2}\right)\cdot \frac{c_2}{c_1}\,,\\
    j_{2\to1} &\simeq -\kappa_0' \cdot \left(\rho_{0,2}+\partial_x\,\rho_{0,2}\cdot \delta s_2\cdot \frac{C_d}{2}\right)\,,
\end{align*}
where
\begin{equation*}
    \kappa_0' = \sqrt{\frac{D_2}{2d\cdot\delta t}}\cdot\frac{1}{C_d}.
\end{equation*}
Therefore, the net particle density flux is
\begin{align} \notag
    j &= j_{1\to2} + j_{2\to1} \\ \label{eq:alt-flux-avg-density}
      &\simeq \kappa_0' \cdot\left[\left(\rho_{0,1}\cdot\frac{c_2}{c_1}-\rho_{0,2}\right) - \frac{C_d}{2}\left( \partial_x\,\rho_{0,1}\cdot \delta s_1 \cdot \frac{c_2}{c_1}+\partial_x\,\rho_{0,2}\cdot \delta s_2 \right)\right]\,,
\end{align}
which indicates the equilibrium condition at $t\to\infty$ limit:
\begin{equation*}
    j\to0\,,\quad \partial_x\,\rho_{0,1}\to0\,,\quad \partial_x\,\rho_{0,2}\to0\,,\quad \frac{\rho_{0,1}}{\rho_{0,2}}\to\frac{c_1}{c_2}\,.
\end{equation*}

Considering the ratio of spin concentrations, we have the net particle density flux
\begin{align} \label{eq:alt-flux-density}
    j &= \kappa'\cdot\left(\rho_{0,1}\cdot\frac{c_2}{c_1} - \rho_{0,2}\right)\\ \label{eq:alt-flux-grad}
      &= -D_1\cdot\partial_x\,\rho_{0,1} = -D_2\cdot \partial_2\,\rho_{0,2}\,,
\end{align}
where $\kappa'$ is the effective permeability due to the finite time-step.
Substituting Eqs.~(\ref{eq:alt-flux-avg-density}) and (\ref{eq:alt-flux-grad}) into \eqref{eq:alt-flux-density} yields
\begin{equation} \label{eq:alt-kappa-correction}
    \kappa' = \frac{\kappa_0'}{1-\beta} \propto \frac{1}{\sqrt{\delta t}}\,,
\end{equation}
where
\begin{align*}
    \beta &= \frac{1}{2}\left(\frac{c_2\sqrt{D_2}}{c_1\sqrt{D_1}}+1\right)\\
          &= \frac{P_{1\to2}+P_{2\to1}}{2}\,.
\end{align*}
Interestingly, similar to $\alpha$ in \eqref{eq:kappa-factor-prob} and $\alpha_\lambda$ in \eqref{eq:gen-kappa-factor-prob}, the correction factor $\beta \in [0.5,1]$ is also the permeation probability averaged for both directions.

For an infinitely small time-step ($\delta t\to0$), the effective permeability is infinitely large ($\kappa'\to\infty$), as predicted by \eqref{eq:alt-kappa-correction}. In this case, the finite density flux $j$ in \eqref{eq:alt-flux-density} indicates no particle density transition over the membrane, i.e. $\rho_{0,1}/\rho_{0,2}\to c_1/c_2$.
Similarly, when $c_1\sqrt{D_1}/c_2\sqrt{D_2}=1$ in \eqref{eq:alt-perm-prob1-2}, the $P_{1\to2}=1$ leads to an infinitely large $\kappa'$, based on \eqref{eq:alt-kappa-correction}, and $\rho_1/\rho_2\to c_1/c_2$ due to the finite density flux $j$ in \eqref{eq:alt-flux-density}.

In contrast, when $P_{1\to2}<1$, a finite time-step $\delta t$ results in an extra effective permeability $\kappa'$, hindering the permeation through membranes. To reduce this unwanted effect, the applied time step needs to be small. For example, considering a multi-compartmental system in $1d$, the size and intrinsic diffusivity in the $i$-th compartment are $l_i$ and $D_i$. Then we can ignore the hindrance through membranes caused by $\kappa'$, if the time-step is sufficiently small, such that
\begin{equation*}
    \kappa'\gg\text{max}\left(\frac{D_i}{l_i}\right)\,,
\end{equation*}
where $D_i/l_i$ is the intrinsic permeability of the $i$-th compartment \citep{novikov2011rpbm}. 
However, in $2d$ and $3d$, the compartment length scale $l_i$ could be ill-defined, complicating the choice of time step $\delta t$.

\section{Estimation of the undulation amplitude and wavelength from a 3\textit{d} axonal skeleton}
\label{sec:app-undulation}
\renewcommand{\theequation}{D.\arabic{equation}}
\setcounter{equation}{0}
\renewcommand{\thefigure}{D.\arabic{figure}}
\setcounter{figure}{0}

In this Appendix, we will introduce how to estimate the length scale of undulation amplitude and wavelength based on a 3{\it d} axonal skeleton. 

Considering an axonal skeleton aligned to its main axis (z-axis), the skeleton can be quantified as ${\bf l} = {\bf w} + z \hat{\bf z}$, with $\bf w$ the deviation of the skeleton from the main axis. We can decompose the axonal skeleton with multiple harmonics \citep{lee2020radial}:
\begin{equation} \notag
    {\bf w} = \sum_n w_{xn} \cos(k_n z + \phi_{xn}) \, \hat{\bf x} + w_{yn} \cos(k_n z+\phi_{yn}) \, \hat{\bf y}\,,
\end{equation}
where, for the $n$-th harmonic ($n\in\mathbb{N}$), $w_{xn}$ and $w_{yn}$ are the undulation amplitudes along x- and y-axes, and $\phi_{xn}$ and $\phi_{yn}$ are the phases. Here we focus on the case of $k_n=n\cdot2\pi/L_z$ (with $L_z$ the axonal length along $z$) such that all harmonics are orthogonal. This allows us to define a Euclidean measure of the undulation amplitude, given by
\begin{equation} \notag
    w_0 \equiv \sqrt{\sum_n w_{0n}^2}\,,\quad w_{0n}\equiv \sqrt{w_{xn}^2+w_{yn}^2}\,.
\end{equation}
The values of $w_{xn}$ and $w_{yn}$ can be estimated by applying discrete cosine transform to $\bf w$, yielding a length scale of the undulation amplitude $\sim w_0$.

Furthermore, the projection factor transverse to the main direction has been shown to be \citep{lee2020radial}
\begin{equation} \label{eq:sin2-nh}
    \langle \sin^2\delta\theta \rangle_z \simeq \frac{2}{\xi^2}\sum_n \left( \frac{\pi w_{0n}}{\lambda_n}\right)^2\,, \quad \xi\simeq 1+\sum_n \left(\frac{\pi w_{0n}}{\lambda_n}\right)^2 \gtrsim 1\,,
\end{equation}
where $\delta \theta = \delta\theta(z)$ is the angle between the individual axon's skeleton segment at $z$ and its main axis, $\langle...\rangle_z$ is the average along each axon's main axis, and $\lambda_n = 2\pi/k_n$ is the corresponding wavelength. To have a rough estimate of the undulation wavelength, we impose a simplified 1-harmonic model ($n\leq 1$) to \eqref{eq:sin2-nh}, leading to
\begin{equation} \label{lambda-w0}
    \lambda \simeq \pi w_0 \sqrt{\frac{2}{\langle \sin^2\delta \theta\rangle_z}}\,,
\end{equation}
where we approximate $\xi\simeq 1$ due to the small $w_{0}/\lambda\sim 0.02$ in realistic axons of a mouse brain \citep{lee2020radial}.

Note that Eq.~(\ref{lambda-w0}) should not generally be interpreted as a proportionality relation between the undulation wavelength and its amplitude, since each axon can have its own dispersion $\delta \theta(z)$. However, empirically, this proportionality seems to be valid based on the high correlation between these two geometric quantities in \figref{fig:ad-rd}d. In other words, the slope 
$\text{d}w_0/\text{d}\lambda \simeq \langle\sin^2\delta\theta\rangle_z^{1/2}/\left(\pi\sqrt{2}\right)$ seems to be sufficiently axon-independent. 
From the estimated slope in that plot, we find the {\it intra-axon} undulation dispersion 
$\sqrt{\langle\delta\theta^2\rangle} \simeq \langle\sin^2\delta \theta\rangle_z^{1/2} \approx 10^\circ$, which is about a factor of 2 smaller than the {\it inter-axon} fiber orientation dispersion $\sqrt{\langle\bar{\theta}^2\rangle}\simeq 22^\circ$ estimated from histology \citep{ronen2014cc,lee2019em}. By using the Rodrigues' rotation formula and the small angle approximation, it is straightforward to combine the two contributions and estimate the overall dispersion angle $\sqrt{\langle \theta^2\rangle}\simeq\sqrt{\langle\bar{\theta}^2\rangle + \langle\delta\theta^2\rangle}\approx 24^\circ$. This means that the ``Standard Model" estimates of dispersion  \citep{novikov2018rotinv,dhital2019axial} is dominated by the inter-axonal contribution, in agreement with the Standard Model assumptions.

\newpage
\bibliographystyle{elsarticle-harv}
\bibliography{reference}

\cleardoublepage
\setcounter{figure}{0}
\setcounter{page}{1}
\renewcommand\thefigure{S.\arabic{figure}}%


\end{document}